\documentclass[12pt]{article}

\usepackage{amsmath, amssymb, amsthm}

\usepackage{newtxtext, newtxmath}
\usepackage[T1]{fontenc}
\usepackage[utf8]{inputenc}
\usepackage{setspace}
\usepackage{microtype}

\usepackage[english]{babel}
\usepackage{csquotes}

\usepackage[margin=1in]{geometry}

\usepackage{booktabs}
\usepackage[hang, small, labelfont=bf, textfont=it, up]{caption}
\usepackage{enumitem}
\setlist[itemize]{noitemsep}
\usepackage{graphicx}
\usepackage{threeparttable}
\DeclareCaptionLabelFormat{AppendixTables}{A.#2}

\usepackage{abstract}

\usepackage{titlesec}
\titleformat*{\section}{\large\bfseries}
\titleformat*{\subsection}{\bfseries}

\usepackage[usenames, dvipsnames]{xcolor}
\definecolor{darkred}{rgb}{0.55, 0.0, 0.0}

\usepackage[hidelinks]{hyperref}
\hypersetup{colorlinks=true, linkcolor=darkred, citecolor=darkred, urlcolor=darkred}

\usepackage[authordate, backend=biber]{biblatex-chicago}
\bibliography{bibliography.bib}

\usepackage{titling}
\pretitle{\begin{center}\Large\bfseries}
\posttitle{\end{center}}
\preauthor{\begin{center}\small\begin{tabular}[t]{c}}
\postauthor{\end{tabular}\end{center}}

\title{Fixed Effects Binary Choice Models: Estimation and Inference with Long Panels\thanks{We thank André Romahn, Joel Stiebale, and Thomas Zylkin for helpful comments. A previous version of this article circulated under the title ``Binary Choice Models with High-Dimensional Individual and Time Fixed Effects''.}}
\author{
	\textsc{Daniel Czarnowske}\thanks{
		Heinrich-Heine-Universität Düsseldorf, Universitätsstr. 1, 40225 Düsseldorf, Germany, e-mail: \texttt{\href{mailto:daniel.czarnowske@hhu.de}{daniel.czarnowske@hhu.de}}
	}
	\and
	\textsc{Amrei Stammann}\thanks{
		Ruhr-Universität Bochum, Universitätsstr. 150, 44801 Bochum, Germany, e-mail: \texttt{\href{mailto:amrei.stammann@rub.de}{amrei.stammann@hhu.de}}
	}
}
\date{\small\today}

\DeclareMathOperator{\aapprox}{\overset{a}{\sim}\,}

\DeclareMathOperator{\argmax}{\arg\,\max\,}
\DeclareMathOperator{\diag}{\text{diag}}
\DeclareMathOperator{\eye}{\mathbf{1}}
\DeclareMathOperator{\ind}{\mathbf{1}}
\DeclareMathOperator{\iid}{\text{iid.}\,}
\DeclareMathOperator{\plim}{\text{plim}\,}

\DeclareMathOperator{\MX}{\mathbb{M}}
\DeclareMathOperator{\PX}{\mathbb{P}}
\DeclareMathOperator{\N}{\mathcal{N}}

\newtheorem*{definition}{Definition}

\begin{document}

\maketitle
\renewcommand{\abstractname}{\vspace{-5em}}
\begin{abstract}
\noindent Empirical economists are often deterred from the application of fixed effects binary choice models mainly for two reasons: the incidental parameter problem and the computational challenge even in moderately large panels.  Using the example of binary choice models with individual and time fixed effects, we show how both issues can be alleviated by combining asymptotic bias corrections with computational advances. Because unbalancedness is often encountered in applied work, we investigate its consequences on the finite sample properties of various (bias corrected) estimators.  In simulation experiments we find that analytical bias corrections perform particularly well, whereas split-panel jackknife estimators can be severely biased in unbalanced panels.\\
\vfill
\end{abstract}
\clearpage
\onehalfspacing

\section{Introduction}
Empirical analyses explaining binary outcomes, such as labour force participation or exporting decisions, are quite common in economics. The increasing number and availability of large and long panel data sets offers several advantages to researchers compared to pure cross-sections or time series  (see chapter 1.2 in \cite{b2013} and \cite{h2014} for a comprehensive list of advantages). Maybe the most important advantage is that they allow to control for different sources of unobserved heterogeneity. In panels it is natural to account for unobserved individual and time specific effects simultaneously, so-called two-way fixed effect models. The corresponding estimators treat the unobserved effects as additional parameters to be estimated and thus allow for unrestricted correlation patterns between the explanatory variables and the unobserved effects. As the researcher does not have to make any distributional assumptions about the unobserved heterogeneity, these models are very flexible and a natural candidate for many empirical applications.

In the early stage of panel data econometrics, panels consisted of relatively few observations per individual. Consequently, when deriving asymptotic properties of estimators, it is very often assumed that the number of individuals ($N$) grows and the number of points in time ($T$) is held fixed. Under this asymptotic framework, non-linear fixed effects estimators are inconsistent, known as the incidental parameter problem (\textit{IPP}) first mentioned by \textcite{ns1948}. This strand of literature is therefore particularly interested in deriving fixed $T$ consistent estimators. For instance, so-called conditional logit estimators have been proposed for static and dynamic binary choice models with individual fixed effects (see \cite{r1960}, \cite{a1970}, \cite{c1980}, and \cite{hk2000}). However, it is not possible to derive fixed $T$ consistent fixed effects estimators for all kind of models, e. g. the probit model. Another drawback of all conditional logit estimators is that they preclude the estimation of partial effects, which are of great interest in economics (see \cite{ah2007}, and \cite{fw2018a}).

For these reasons, among others, and further motivated by the seminal work of \textcite{pm1999} and the rising availability of comprehensive panel data, a growing literature now focuses on large $N$ and $T$ asymptotics. The beauty of this asymptotic framework is that the inconsistency problem of the \textit{IPP} can be turned into an asymptotic bias problem that can be corrected. More precisely, \textcite{hk2002} were the first to exploit this asymptotic framework and to propose a bias-corrected estimator for dynamic linear panel models with individual fixed effects to address the inference problem induced by the \textcite{n1981} bias. \textcite{hm2006} show that the same bias correction is also applicable if the model includes additional time fixed effects. In the meantime, several bias-corrected estimators for non-linear models have been proposed (see among others \cite{l2002}, \cite{w2002}, \cite{hn2004}, \cite{c2007}, \cite{f2009}, \cite{bh2009}, \cite{dj2015}, \cite{fw2016}, and \cite{ks2016}). A remarkable difference compared to estimators for linear models is that the inclusion of time fixed effects leads to an additional bias as shown by \cite{fw2016}. We refer the interested reader to \textcite{ah2007} and \textcite{fw2018a} for comprehensive overviews.

Another apparent challenge that discourages researchers from using non-linear fixed effects models is the computational burden associated with the estimation. This problem is especially severe when the model specification leads to high-dimensional fixed effects, which is already the case when $N$ or $T$ are large. If only one of the panel dimensions is large, the algorithm of \textcite{g2004} can significantly reduce the computational burden. If both dimensions are large algorithms like \textcite{gp2010} and \textcite{s2018} can be used. From a practical point of view, however, it is not so obvious how these algorithms can be combined with analytical bias corrections like the one of \textcite{fw2016}.

In this article, we offer new insights that facilitate and validate the usage of different (bias-corrected) estimators for binary choice models with two-way fixed effects in empirical research. First, we show how to address the computational obstacles that often prevent the application of bias corrections. Second, we extend the simulation experiments of \textcite{fw2016} by several aspects to gain deeper insights into the statistical properties of different estimators. More specifically, we analyse further analytical and split-panel jackknife bias-corrected estimators that were proposed but not studied by the authors. We also consider alternative estimators for average partial effects based on linear fixed effects models. These models are often used in empirical research to avoid the above mentioned pitfalls of non-linear ones. Because many real world panel data sets are unbalanced, our analysis also considers different patterns of randomly missing data. This aspect has so far received little attention in the literature, but is relevant for many empirical applications. Overall, we find that analytical bias corrections are preferable to split-panel jackknife approaches. In general, the latter show higher distortion, lower coverage, and are less robust to different patterns of randomly missing data. Third, we provide an illustrative example using an unbalanced panel from the German Socio-Economic Panel (see \cite{wfs2007}) to investigate the inter-temporal labour force participation of 6,241 women between 1984 and 2013. Inspired by \textcite{h1999} we estimate a dynamic fixed effects probit model where bias corrections are required to deal with the inference problem. Finally, we offer the analytical bias correction of \textcite{fw2016} in our \texttt{R} package \textit{alpaca} to encourage its application.\footnote{Until now, the analytical bias correction proposed by \textcite{fw2016} is only provided in a \textit{Stata} routine by \textcite{cfw2017}, which is not designed for long panels.}

The remainder of this article is organized as follows. In Section \ref{sec:bc} we introduce the model, various bias corrections, and algorithms for handling panels with large $N$ and $T$. In Section \ref{sec:sim} we provide results of extensive simulation experiments. In Section \ref{sec:illustration} we apply different bias-corrected estimators to an empirical example from labour economics. Finally, we give some concluding remarks in Section \ref{sec:conclusion}.

Throughout this article, we follow conventional notation: scalars are represented in standard type, vectors and matrices in boldface, and all vectors are column vectors.

\section{Bias Corrections for Fixed Effects Binary Choice Models}
\label{sec:bc}

\subsection{Model, Assumptions, and the Inference Problem}
The fixed effects binary choice model examined in this article can be derived from a latent variable model with two additive unobserved effects. Let
\begin{equation*}
y_{it}^{\ast} = \mathbf{x}_{it}^{\prime} \boldsymbol{\beta} + \alpha_{i} + \gamma_{t} + e_{it} \, ,
\end{equation*}
be the latent variable, where $i$ and $t$ are individual and time specific indexes.\footnote{Without loss of generality, $i$ and $t$ could also be indexes of networked economic activities like trade between countries.} However, instead of the latent variable, we only observe $y_{it} = 1$ if $y_{it}^{\ast} \geq 0$ and $y_{it} = 0$ otherwise. To allow for missing data, we define the following sets: $\mathcal{S}$ is a subset of $\{(i, t) \mid i \in \{1, \ldots, N\} \wedge t \in \{1, \ldots, T\}\}$ containing all observed pairs of indexes and $\mathcal{S}_{t} = \{i \mid (i, t) \in \mathcal{S}\}$ and $\mathcal{S}_{i} = \{t \mid (i, t) \in \mathcal{S}\}$ are subsets of $\mathcal{S}$ containing all indexes of individuals observed for a period $t$ and points in time observed for an individual $i$. $N$ and $T$ are the number of individuals and points in time and $n = \lvert \mathcal{S} \rvert$ is the sample size. Furthermore, $\mathbf{x}_{it}$ is the $it$-th row of the regressor matrix $\mathbf{X}$ and a $J$-dimensional vector of possibly predetermined explanatory variables, $\boldsymbol{\beta}$ are the corresponding structural parameters, $\alpha_{i}$ and $\gamma_{t}$ are incidental parameters that capture unobserved individual and time specific effects, and $e_{it}$ is an idiosyncratic error term assumed to be mean zero and independent of $\mathbf{X}_{i}^{\star} = (\mathbf{x}_{is})_{s \leq t \in \mathcal{S}_{i}}$ and $\boldsymbol{\pi} = (\boldsymbol{\alpha}, \boldsymbol{\gamma})$, where $\boldsymbol{\alpha} = (\alpha_{1}, \ldots, \alpha_{N})$ and $\boldsymbol{\gamma} = (\gamma_{1}, \ldots, \gamma_{T})$. Note that the imposed exogeneity assumption is milder than the often assumed strict exogeneity condition, where $e_{it}$ is independent of $\mathbf{X}_{i} = (\mathbf{x}_{is})_{s \in \mathcal{S}_{i}}$ instead of $\mathbf{X}_{i}^{\star}$. This stricter assumption is often too restrictive in the context of panel data, because it is violated if future realizations of a regressor are affected by the outcome variable, which for instance also precludes lagged dependent variables as regressors.  In the presence of missing data, we have to additionally assume that conditional on $\mathbf{X}_{i}^{\star}$ and $\boldsymbol{\pi}$ the observations are missing at random.

Assuming a certain distribution for $e_{it}$ allows us to use the principle of maximum likelihood to derive a parametric estimator for fixed effects binary choice models. Let
\begin{equation*}
	l_{it}(\boldsymbol{\beta}, \alpha_{i}, \gamma_{t}) = y_{it} \log(F_{it}) + (1 - y_{it}) \log(1 - F_{it})
\end{equation*}
be the log-likelihood contribution of individual $i$ at time $t$, where $F_{it}$ is the cumulative distribution function of $e_{it}$ evaluated at the linear index $\eta_{it} = \mathbf{x}_{it}^{\prime} \boldsymbol{\beta} + \alpha_{i} + \gamma_{t}$. Common choices for $F_{it}$ are the standard normal, the logistic, and the complementary log-log distribution. The corresponding maximum likelihood estimator is
\begin{equation*}
	\hat{\boldsymbol{\theta}} = (\hat{\boldsymbol{\beta}}, \hat{\boldsymbol{\pi}}) = \underset{\boldsymbol{\beta} \in \mathbb{R}^{J}, \boldsymbol{\alpha} \in \mathbb{R}^{N}, \boldsymbol{\gamma} \in \mathbb{R}^{T}}{\argmax} \sum_{(i, t) \in \mathcal{S}} l_{it}(\boldsymbol{\beta}, \alpha_{i}, \gamma_{t}) \, .
\end{equation*}

Although \textcite{fw2016} show consistency of $\hat{\boldsymbol{\beta}}$ under asymptotics where $N$ and $T$ grow at the same rate ($\plim_{N, T \rightarrow \infty} \hat{\boldsymbol{\beta}} = \boldsymbol{\beta}$), they also expose an asymptotic bias in the limiting distribution of the estimator with some severe consequences for inference. To get a better understanding of this specific inference problem, we briefly summarize the key findings of \textcite{fw2016} for binary choice models and combine them with their conjecture about unbalanced panels stated in \textcite{fw2018a}. 

Under asymptotic sequences where $N, T \rightarrow \infty$, $N / T \rightarrow \kappa^{2}$, and $0 < \kappa < \infty$ plus certain regularity conditions, like additively separable unobserved effects and concavity of the objective function, an asymptotic approximation to the limiting distribution of $\hat{\boldsymbol{\beta}}$ is given by
\begin{equation*}
	\hat{\boldsymbol{\beta}} \aapprox \N (\boldsymbol{\beta} + \overline{T}^{- 1} \mathbf{B}^{\beta} + \overline{N}^{- 1} \mathbf{C}^{\beta}, \; \mathbf{V}^{\beta}) \, ,
\end{equation*}  
where $\overline{N} = n / T$ and $\overline{T} = n / N$ are the average number of individuals and points in time, $\mathbf{B}^{\beta}$ and $\mathbf{C}^{\beta}$ are the leading terms of the asymptotic bias $\mathbf{b}^{\beta} = \overline{T}^{- 1} \mathbf{B}^{\beta} + \overline{N}^{- 1} \mathbf{C}^{\beta}$ stemming from the inclusion of individual and time specific fixed effects, and $\mathbf{V}^{\beta}$ is the asymptotic covariance matrix. Due the asymptotic bias, the approximated limiting distribution is not correctly centred at $\boldsymbol{\beta}$, which means that, even if $n$ is large, confidence intervals constructed around any $\hat{\boldsymbol{\beta}}$ might not cover the true value of the corresponding parameter with probability close to the desired nominal level. However, this inference problem can be corrected by forming suitable estimators for $\mathbf{b}^{\beta}$ that can be subtracted from $\hat{\boldsymbol{\beta}}$.

Most researchers are not directly interested in the structural parameters, but rather in average partial effects. Let
\begin{equation*}
	\boldsymbol{\Delta}_{itj} = \begin{cases}
		\beta_{j} \partial_{\eta} F_{it} & (\text{continous regressor})\\
		F_{it}\rvert_{x_{itj} = 1} - F_{it}\rvert_{x_{itj} = 0} & (\text{binary regressor})
	\end{cases}
\end{equation*}
denote the partial effect of a change in $x_{itj}$, where $x_{itj}$ is the $j$-th element in $\mathbf{x}_{it}$, $\partial_{\eta} F_{it}$ is the first-order partial derivative of $F_{it}$ with respect to $\eta_{it}$, and $F_{it}\rvert_{x_{itj} = k}$ indicates that $x_{itj}$ in the linear index is replaced by $k$.\footnote{For simplicity, we ignore the possibility of more complicated  functional forms in the linear index, e. g. polynomials.} The average partial effects are then given by $\boldsymbol{\delta} = (\delta_{1}, \ldots, \delta_{J})$, where $\delta_{j} = n^{- 1} \sum_{(i, t) \in \mathcal{S}} \boldsymbol{\Delta}_{itj}$. Under some additional sampling and moment conditions, an asymptotic approximation to the limiting distribution of the estimator of the average partial effects is given by
\begin{equation*}
	\hat{\boldsymbol{\delta}} \aapprox \N(\boldsymbol{\delta} + \overline{T}^{-1} \mathbf{B}^{\delta} + \overline{N}^{- 1} \mathbf{C}^{\delta}, \; \mathbf{V}^{\delta}) \, ,
\end{equation*}
where $\mathbf{V}^{\delta}$ is the asymptotic covariance matrix. Again, $\mathbf{B}^{\delta}$ and $\mathbf{C}^{\delta}$ are the leading terms of the asymptotic bias $\mathbf{b}^{\delta} = \overline{T}^{- 1} \mathbf{B}^{\delta} + \overline{N}^{- 1} \mathbf{C}^{\delta}$ stemming from the inclusion of individual and time specific fixed effects. Thus, as for $\hat{\boldsymbol{\beta}}$, there is an asymptotic bias problem.

In the next subsection we summarize the different bias corrections proposed by \textcite{fw2016}, but using a slightly modified notation. We change their notation for two reasons: First, we are considering panels that are potentially unbalanced, and second, we want to emphasize the link to recent advances in computational econometrics, which allow us to estimate (bias-corrected) binary choice models even when both panel dimensions are large.

\subsection{Asymptotic Bias Corrections}

Before we present the different bias corrections proposed by \textcite{fw2016}, we have to introduce some additional notation. Let $\partial_{z^{p}} G_{it}$ denote the $p$-th order partial derivative of an arbitrary function $G_{it}$ with respect to $z_{it}$. Further, let $\partial_{\eta} l_{it} =  H_{it} (y_{it} -  F_{it})$, $\omega_{it} =  H_{it} \partial_{\eta}  F_{it}$,  $H_{it} = \partial_{\eta}  F_{it} / (F_{it} (1 - F_{it}))$, and $\nu_{it} = \partial_{\eta} l_{it} / \omega_{it}$. We use vector notation to indicate that we collect the different quantities for all observations, e. g. $\boldsymbol{\omega} = (\omega_{it})_{(i, t) \in \mathcal{S}}$. Finally we define the residual projection $\MX = \eye_{n} - \PX = \eye_{n} - \mathbf{D}(\mathbf{D}^{\prime} \boldsymbol{\Omega} \mathbf{D})^{+} \mathbf{D}^{\prime} \boldsymbol{\Omega}$, where $\eye_{n}$ is an identity matrix of dimension $(n \times n)$, $\mathbf{D}$ is a sparse indicator matrix of dimension $(n \times N + T)$ arising from dummy encoding of individual and time identifiers, $(\cdot)^{+}$ refers to the Moore-Penrose inverse, and $\boldsymbol{\Omega}$ is a positive definite diagonal weighting matrix with $\diag(\boldsymbol{\Omega}) = \boldsymbol{\omega}$. The corresponding sample analogues are indicated by a hat. For clarification, we refer to $\hat{\eta}_{it} = \mathbf{x}_{it}^{\prime} \hat{\boldsymbol{\beta}} + \hat{\alpha}_{i} + \hat{\gamma}_{t}$ as the sample analogue of $\eta_{it}$. Table \ref{tab:expressions} contains explicit expressions for distributions and the corresponding derivatives of frequently used binary choice models.
\begin{table}[!htbp]
	\centering
	\caption{\label{tab:expressions}Distributions and their Derivatives.}
	\begin{threeparttable}
		\begin{tabular}{@{}lccc@{}}
			\toprule
			&                                  Logit                                   &                Probit                 &                            Complementary Log-Log                            \\
			\midrule
			$F_{it}$                     &                      $(1 + \exp(- \eta_{it}))^{-1}$                      &           $\Phi(\eta_{it})$           &                        $1 - \exp(- \exp(\eta_{it}))$                        \\
			$\partial_{\eta} F_{it}$     &                          $F_{it} (1 - F_{it})$                           &           $\phi(\eta_{it})$           &                     $\exp(\eta_{it} - \exp(\eta_{it}))$                     \\
			$\partial_{\eta^{2}} F_{it}$ &                 $\partial_{\eta} F_{it} (1 - 2 F_{it})$                  &     $- \eta_{it} \phi(\eta_{it})$     &               $\partial_{\eta} F_{it} (1 - \exp(\eta_{it}))$                \\
			$\partial_{\eta^{3}} F_{it}$ & $\partial_{\eta} F_{it} ((1 - 2 F_{it})^{2} - 2 \partial_{\eta} F_{it})$ & $(\eta_{it}^{2} - 1) \phi(\eta_{it})$ & $\partial_{\eta^{2}} F_{it} (2 - \exp(\eta_{it})) - \partial_{\eta} F_{it}$ \\
			\bottomrule
		\end{tabular}
		\begin{tablenotes}
			\footnotesize
			\item\emph{Note:} $\Phi(\cdot)$ and $\phi(\cdot)$ are the cumulative distribution and probability density function of the standard normal distribution.
		\end{tablenotes}
	\end{threeparttable}
\end{table}

\textcite{fw2016} distinguish between two types of bias corrections: analytical and split-panel jackknife. The latter exploits the relation between sample size and bias to form a non-parametric estimator of the asymptotic bias and is an extension of \textcite{dj2015}, whereas the former relies on explicit expressions derived from asymptotic expansions.\footnote{The idea to reduce bias using jackknife techniques originates from \textcite{q1949, q1956}. In the context of dynamic models they were first mentioned by \textcite{h2002}.} A bias-corrected estimator for $\boldsymbol{\beta}$ is
\begin{equation*}
	\tilde{\boldsymbol{\beta}} = \hat{\boldsymbol{\beta}} - \hat{\mathbf{b}}^{\beta} \, ,
\end{equation*}
where $\hat{\mathbf{b}}^{\beta}$ is an estimator of the asymptotic bias such that $\tilde{\boldsymbol{\beta}} \aapprox \N (\boldsymbol{\beta}, \mathbf{V}^{\beta})$.

We start with an analytical bias correction at the level of the estimator. An explicit expression for an estimator of the asymptotic bias is
\begin{equation*}
	\hat{\mathbf{b}}_{\text{abc}}^{\beta} = \widehat{\mathbf{W}}^{- 1} (\widehat{\mathbf{B}}^{\beta} + \widehat{\mathbf{C}}^{\beta}) \, ,
\end{equation*}
where
\begin{eqnarray}
	\widehat{\mathbf{B}}^{\beta} &=& - \frac{1}{2} \sum_{i = 1}^{N} \frac{\sum_{t \in \mathcal{S}_{i}} \widehat{H}_{it} \partial_{\eta^{2}} \widehat{F}_{it} (\widehat{\MX}\mathbf{X})_{it} + 2 \sum_{l = 1}^{L} \tau_{i}(l)  \sum_{t > l \in \mathcal{S}_{i}} \partial_{\eta} \hat{l}_{it-l} \hat{\omega}_{it} (\widehat{\MX}\mathbf{X})_{it}}{\sum_{t \in \mathcal{S}_{i}} \hat{\omega}_{it}}  \, , \nonumber \\
	\widehat{\mathbf{C}}^{\beta} &=& - \frac{1}{2} \sum_{t = 1}^{T} \frac{\sum_{i \in \mathcal{S}_{t}} \widehat{H}_{it} \partial_{\eta^{2}} \widehat{F}_{it} (\widehat{\MX}\mathbf{X})_{it}}{\sum_{i \in \mathcal{S}_{t}} \hat{\omega}_{it}} \, , \nonumber \\
	\widehat{\mathbf{W}} &=& \sum_{(i, t) \in \mathcal{S}} \hat{\omega}_{it} (\widehat{\MX}\mathbf{X})_{it} (\widehat{\MX}\mathbf{X})_{it}^{\prime} \, , \nonumber 
\end{eqnarray}
$L$ is the bandwidth parameter of the truncated spectral density estimator suggested by \textcite{hk2007}, and $\tau_{i}(l)  = \lvert \mathcal{S}_{i} \rvert / (\lvert \mathcal{S}_{i} \rvert - l)$ is a finite sample adjustment proposed by \textcite{fw2016}. The corresponding estimator of the asymptotic covariance is $\widehat{\mathbf{V}}^{\beta} = \widehat{\mathbf{W}}^{- 1}$. By making the stronger strict exogeneity assumption, we can set $L = 0$ and drop the second term in $\widehat{\mathbf{B}}^{\beta}$, so that the expressions for $\widehat{\mathbf{B}}^{\beta}$ and $\widehat{\mathbf{C}}^{\beta}$ become identical except for the indexes.\footnote{The second term in $\widehat{\mathbf{B}}^{\beta}$ can be interpreted as an estimator for a \textcite{n1981}-type bias.} However, this stronger assumption is difficult to motivate in practice. Therefore, \textcite{fw2016, fw2018a} recommend to check the sensitivity of the estimates using different values of $L \in \{0, \ldots ,4\}$.

To possibly further improve its finite sample properties, the analytical bias correction can be further iterated. More precisely, we start with an initial $\tilde{\boldsymbol{\beta}}$, afterwards we recompute $\hat{\mathbf{b}}^{\text{abc}}$, update $\tilde{\boldsymbol{\beta}}$, and repeat this procedure a finite number of times. \textcite{ah2007} refer to this approach as infinitely repeated analytical bias correction.

Next we describe how the split-panel jackknife can be used to form estimators of the asymptotic bias. The idea is to split the panel into sub panels and use them to construct an estimator of the asymptotic bias from different sub panel estimators. We consider two different splitting strategies: the first strategy (\textit{SPJ1}) is described in \textcite{fw2016} and the second one (\textit{SPJ2}) in \textcite{cfw2017}. Let 
\begin{equation*}
	\hat{\mathbf{b}}_{\text{spj1}}^{\beta} = \hat{\boldsymbol{\beta}}^{N} + \hat{\boldsymbol{\beta}}^{T} - 2 \hat{\boldsymbol{\beta}} \quad \text{and} \quad \hat{\mathbf{b}}_{\text{spj2}}^{\beta} = \hat{\boldsymbol{\beta}}^{NT} - \hat{\boldsymbol{\beta}}
\end{equation*}
be estimators of the asymptotic bias, where
\begin{eqnarray}
	\hat{\boldsymbol{\beta}}^{N\phantom{T}} &=& \frac{1}{2} \big(\hat{\boldsymbol{\beta}}_{\{i \leq \lceil N / 2 \rceil\}} + \hat{\boldsymbol{\beta}}_{\{i \geq \lfloor N / 2 + 1 \rfloor\}}\big) \, , \nonumber \\
	\hat{\boldsymbol{\beta}}^{T\phantom{N}} &=& \frac{1}{2} \big(\hat{\boldsymbol{\beta}}_{\{t \leq \lceil T / 2 \rceil\}} + \hat{\boldsymbol{\beta}}_{\{t \geq \lfloor T / 2 + 1 \rfloor\}}\big) \, , \nonumber \\
	\hat{\boldsymbol{\beta}}^{NT} &=& \frac{1}{4} \big(\hat{\boldsymbol{\beta}}_{\{i \leq \lceil N / 2 \rceil \wedge t \leq \lceil T / 2 \rceil\}} + \hat{\boldsymbol{\beta}}_{\{i \leq \lceil N / 2 \rceil \wedge t \geq \lfloor T / 2 + 1 \rfloor\}} + \nonumber \\
	&&\phantom{\frac{1}{4} \big(}\hat{\boldsymbol{\beta}}_{\{i \geq \lfloor N / 2 + 1 \rfloor \wedge t \leq \lceil T / 2 \rceil\}} + \hat{\boldsymbol{\beta}}_{\{i \geq \lfloor N / 2 + 1 \rfloor \wedge t \geq \lfloor T / 2 + 1 \rfloor\}}\big) \, , \nonumber
\end{eqnarray}
$\lceil \cdot \rceil$ and $\lfloor \cdot \rfloor$ are floor and ceiling functions, and the subscript in curly brackets indicates the condition to construct the corresponding sub panel. For clarification, $\{i \leq \lceil N / 2 \rceil \wedge t \leq \lceil T / 2 \rceil\}$ means that the corresponding sub panel only contains the first half of all individuals in the first half of the observation period. In the presence of missing data, we follow the suggestion of \textcite{fw2018a} and ignore the attrition process. Note that, contrary to analytical bias corrections, the split-panel jackknife requires an additional unconditional homogeneity assumption (see assumption 4.3 in \cite{fw2016}). For instance, this condition rules out trends or structural breaks in the explanatory variables. Further note that both splitting strategies can lead to overlapping sub panels that introduce an additional variance inflation as pointed out by \textcite{dj2015}.\footnote{\textcite{dj2015} show how to construct non-overlapping sub panels.}

The analytical bias corrections can also be applied at the level of score. The corresponding bias-corrected estimator is the solution to the following system of equations:
\begin{equation*}
	(\widehat{\MX}\mathbf{X}(\boldsymbol{\beta}))^{\prime} \widehat{\boldsymbol{\Omega}}(\boldsymbol{\beta})  \hat{\boldsymbol{\nu}}(\boldsymbol{\beta}) = \widehat{\mathbf{B}}^{\beta} + \widehat{\mathbf{C}}^{\beta} \, .
\end{equation*}
\textcite{fw2016}  also suggest a continuously updated score correction where $\widehat{\mathbf{B}}^{\beta}$ and $\widehat{\mathbf{C}}^{\beta}$ are replaced by $\widehat{\mathbf{B}}^{\beta}(\boldsymbol{\beta})$ and $\widehat{\mathbf{C}}^{\beta}(\boldsymbol{\beta})$ such that $\boldsymbol{\beta}$ and the asymptotic bias are estimated simultaneously.

Finally we describe the bias corrections for the average partial effects. A bias-corrected estimator for $\boldsymbol{\delta}$ is
\begin{equation*}
	\tilde{\boldsymbol{\delta}} = \hat{\boldsymbol{\delta}} - \hat{\mathbf{b}}^{\delta} \, ,
\end{equation*}
where $\hat{\mathbf{b}}^{\delta}$ is an estimator of the asymptotic bias such that $\tilde{\boldsymbol{\delta}} \aapprox \N (\boldsymbol{\delta}, \mathbf{V}^{\delta})$. Again, we can either use explicit expressions or we can use the split-panel jackknife described earlier to construct an estimator of the asymptotic bias. Because the application of the split-panel jackknife is generic and already known from the estimation of $\boldsymbol{\beta}$, we omit it for brevity

In the following we assume that $\hat{\boldsymbol{\Delta}}_{it}$ and $\hat{\boldsymbol{\delta}}$ are constructed from $\tilde{\boldsymbol{\beta}}$ and $\tilde{\boldsymbol{\pi}}$, where
\begin{equation*}
\tilde{\boldsymbol{\pi}} = (\tilde{\boldsymbol{\alpha}}, \tilde{\boldsymbol{\gamma}}) = \underset{\boldsymbol{\alpha} \in \mathbb{R}^{N}, \boldsymbol{\gamma} \in \mathbb{R}^{T}}{\argmax} \sum_{(i, t) \in \mathcal{S}} l_{it}(\tilde{\boldsymbol{\beta}}, \alpha_{i}, \gamma_{t}) \, .
\end{equation*}
An analytical estimator of the asymptotic bias is
\begin{equation*}
	\hat{\mathbf{b}}_{\text{abc}}^{\delta} = n^{- 1} (\widehat{\mathbf{B}}^{\delta} + \widehat{\mathbf{C}}^{\delta}) \, ,
\end{equation*}
where
\begin{eqnarray}
	\widehat{\mathbf{B}}^{\delta} &=& \frac{1}{2} \sum_{i = 1}^{N} \frac{\sum_{t \in \mathcal{S}_{i}} \widehat{H}_{it} \partial_{\eta^{2}} \widehat{F}_{it} (\widehat{\PX}\widehat{\boldsymbol{\Psi}})_{it} + \partial_{\eta^{2}} \hat{\boldsymbol{\Delta}}_{it} - 2 \sum_{l = 1}^{L} \tau_{i}(l)  \sum_{t > l \in \mathcal{S}_{i}} \partial_{\eta} \hat{l}_{it-l} \hat{\omega}_{it} (\widehat{\MX}\widehat{\boldsymbol{\Psi}})_{it}}{\sum_{t \in \mathcal{S}_{i}} \hat{\omega}_{it}} \, , \nonumber \\
	\widehat{\mathbf{C}}^{\delta} &=& \frac{1}{2} \sum_{t = 1}^{T} \frac{\sum_{i \in \mathcal{S}_{t}} \widehat{H}_{it} \partial_{\eta^{2}} \widehat{F}_{it} (\widehat{\PX}\widehat{\boldsymbol{\Psi}})_{it} + \partial_{\eta^{2}} \hat{\boldsymbol{\Delta}}_{it}}{\sum_{i \in \mathcal{S}_{t}} \hat{\omega}_{it}} \, , \nonumber
\end{eqnarray}
and $\widehat{\boldsymbol{\Psi}}_{it} = - \partial_{\eta} \hat{\boldsymbol{\Delta}}_{it} / \hat{\omega}_{it}$. Again, by making the stronger strict exogeneity assumption, we can set $L = 0$ and drop the last term in $\widehat{\mathbf{B}}^{\delta}$. Let $\bar{\hat{\boldsymbol{\Delta}}}_{it} = \hat{\boldsymbol{\Delta}}_{it} - \hat{\boldsymbol{\delta}}$, the corresponding estimator of the asymptotic covariance is
\begin{equation*}
	\widehat{\mathbf{V}}^{\delta} = n^{- 2}\Big(\Big(\sum_{(i, t) \in \mathcal{S}} \bar{\hat{\boldsymbol{\Delta}}}_{it}\Big) \Big(\sum_{(i, t) \in \mathcal{S}} \bar{\hat{\boldsymbol{\Delta}}}_{it}\Big)^{\prime} + \sum_{(i, t) \in \mathcal{S}} \widehat{\boldsymbol{\Gamma}}_{it} \widehat{\boldsymbol{\Gamma}}_{it}^{\prime} + 2 \sum_{i = 1}^{N}\sum_{s > t \in \mathcal{S}_{i}} \bar{\hat{\boldsymbol{\Delta}}}_{it} \widehat{\boldsymbol{\Gamma}}_{is}^{\prime}\Big) \, ,
\end{equation*}
where
\begin{equation*}
	\widehat{\boldsymbol{\Gamma}}_{it} = \Big(\sum_{(i, t) \in \mathcal{S}} \partial_{\beta} \hat{\boldsymbol{\Delta}}_{it} - (\widehat{\PX}\mathbf{X})_{it} \partial_{\eta} \hat{\boldsymbol{\Delta}}_{it}\Big)^{\prime} \widehat{\mathbf{W}}^{- 1} (\widehat{\MX}\mathbf{X})_{it} \partial_{\eta} \hat{l}_{it} - (\widehat{\PX}\widehat{\boldsymbol{\Psi}})_{it} \partial_{\eta} \hat{l}_{it} \, .
\end{equation*}
The first and second term measure the uncertainty caused by the substitution of population by sample means and by the estimation of the structural parameters. The last term is a covariance between both sources of uncertainty that can be dropped if we make the stronger strict exogeneity assumption.\footnote{\label{fn:vcov}The estimator of the asymptotic covariance can be adjusted to take into account additional sampling assumptions with respect to the unobserved effects. For instance, if $\{\alpha_{i}\}_{N}$ and $\{\gamma_{t}\}_{T}$ are assumed to be sequences of independent random variables, where $\alpha_{i} \perp \gamma_{t} \, \forall \, i, t$, the estimator of the asymptotic covariance becomes
\begin{equation*}
	\widehat{\mathbf{V}}^{\delta} = n^{- 2} \sum_{i = 1}^{N} \Big(\sum_{(t, s) \in \mathcal{S}_{i}} \bar{\hat{\boldsymbol{\Delta}}}_{it} \bar{\hat{\boldsymbol{\Delta}}}_{is}^{\prime} + \sum_{(i^{\prime}, t) \in \mathcal{S}} \bar{\hat{\boldsymbol{\Delta}}}_{it} \bar{\hat{\boldsymbol{\Delta}}}_{i^{\prime}t}^{\prime} + \sum_{t \in \mathcal{S}_{i}} \widehat{\boldsymbol{\Gamma}}_{it} \widehat{\boldsymbol{\Gamma}}_{it}^{\prime} + 2 \sum_{s > t \in \mathcal{S}_{i}} \bar{\hat{\boldsymbol{\Delta}}}_{it} \widehat{\boldsymbol{\Gamma}}_{is}^{\prime}\Big)  \, .
\end{equation*}}

So far we learned how to mitigate the inference problem. In the next subsection we show how the computational costs associated with the estimation and the application of bias corrections can be reduced substantially.

\subsection{Feasible Estimation with Long Panel Data}
\label{sec:comp}

In applications where both panel dimensions are large, estimation with standard software quickly becomes very time-consuming or even infeasible. However, recent advances in computational econometrics embed special solvers in the optimization algorithm of the maximum likelihood estimator to address this problem (see \cite{gp2010} and \cite{s2018}). We first summarize the basic idea of \textcite{s2018}, who proposed an algorithm that can be interpreted as a generalization of \textcite{g2004} to more than one fixed effect, and then present an extension that can be used to estimate $\boldsymbol{\pi}$ for a given $\tilde{\boldsymbol{\beta}}$.\footnote{To be more precise, \textcite{s2016} show that the algorithm of \textcite{g2004} can also be derived using the Frisch-Waugh-Lovell theorem. \textcite{s2018} combines the resulting projections with \textcite{h1962}'s method of alternating projections, leading to a very efficient algorithm for any number of fixed effects.} This extension is necessary e. g. for the bias corrections of the average partial effects. Details about the derivation, the algorithms, and an example code are included in the lementary material.

We start with the estimation of $\boldsymbol{\beta}$. Each step of the optimization algorithm involves solving a weighted least squares problem. More precisely, in iteration $r$ we set
\begin{equation*}
	\boldsymbol{\theta}^{[r + 1]} = (\boldsymbol{\beta}^{[r + 1]}, \boldsymbol{\pi}^{[r + 1]}) = (\mathbf{Z}^{\prime} \boldsymbol{\Omega}^{[r]} \mathbf{Z})^{+} \mathbf{Z}^{\prime} \boldsymbol{\Omega}^{[r]} \mathbf{w}^{[r]} \, ,
\end{equation*}
where $\mathbf{Z} = (\mathbf{X}, \mathbf{D})$ and  $\mathbf{w}^{[r]} = \boldsymbol{\nu}^{[r]} + \boldsymbol{\eta}^{[r]}$. Remember, $\boldsymbol{\Omega}^{[r]}$ is a diagonal matrix with $\boldsymbol{\omega}^{[r]} = \diag(\boldsymbol{\Omega}^{[r]})$ and $\boldsymbol{\eta}^{[r]}$ is the collection of linear indexes. Because the rank of $\mathbf{D}$ increases with the sample size, solving the optimization problem quickly becomes infeasible. However we can formulate an alternative weighted least squares problem based on ``demeaned'' variables so that
\begin{equation*}
	\boldsymbol{\beta}^{[r + 1]} = ((\MX\mathbf{X}^{[r]})^{\prime} \boldsymbol{\Omega}^{[r]} \MX\mathbf{X}^{[r]})^{- 1} (\MX\mathbf{X}^{[r]})^{\prime} \boldsymbol{\Omega}^{[r]} \MX\mathbf{w}^{[r]} \, .
\end{equation*}
Afterwards we can use
\begin{equation*}
	\boldsymbol{\eta}^{[r + 1]} = \mathbf{w}^{[r]} - (\MX\mathbf{w}^{[r]} - \MX\mathbf{X}^{[r]} \boldsymbol{\beta}^{[r + 1]})
\end{equation*}
to update $\boldsymbol{\omega}$ and $\mathbf{w}$ for the subsequent iteration. Consequently, $\boldsymbol{\beta}$ are the coefficients obtained by a regression of a weighted two-way demeaned $\mathbf{w}$ on a weighted two-way demeaned $\mathbf{X}$ using $\boldsymbol{\omega}$ as weights. $\boldsymbol{\eta}$ are the corresponding fitted values. Thus we can update $\boldsymbol{\beta}$, $\mathbf{w}$, and $\boldsymbol{\omega}$ in each iteration without explicitly updating the incidental parameters. The practical advantage is that $\boldsymbol{\beta}$ and $\boldsymbol{\eta}$ can be very efficiently computed with any software routine developed for weighted least squares problems with high-dimensional fixed effects.\footnote{Some examples available in popular statistical software are \textit{reghdfe} by \textcite{c2016} for \textit{Stata}, \textit{lfe} by \textcite{g2013a} for \textit{R}, \textit{FixedEffectModels} by Matthieu Gomez for \textit{Julia}, and \textit{pyhdfe} by Jeff Gortmaker for \textit{Python}.}  Note that we can also use the same software routines to compute all terms that depend on $\MX$ and $\PX$, as in some expressions of the bias corrections.

Next, we modify the algorithm of \textcite{s2018} to estimate $\mathbf{D}\boldsymbol{\pi}$ given a fixed $\tilde{\boldsymbol{\beta}}$. Note that we estimate $\mathbf{D}\boldsymbol{\pi}$ instead of $\boldsymbol{\pi}$ as this is simpler and fully sufficient to compute the linear index needed for $F_{it}$ and its derivatives. Again, we start with the weighted least squares problem in iteration $r$:
\begin{equation*}
	\boldsymbol{\pi}^{[r + 1]} = (\mathbf{D}^{\prime} \boldsymbol{\Omega}^{[r]} \mathbf{D})^{+} \mathbf{D}^{\prime} \boldsymbol{\Omega}^{[r]} \tilde{\mathbf{w}}^{[r]} \, ,
\end{equation*}
where $\tilde{\mathbf{w}}^{[r]} = \mathbf{w}^{[r]} - \mathbf{X} \tilde{\boldsymbol{\beta}}$ and $\boldsymbol{\eta}^{[r]} = \mathbf{X}\tilde{\boldsymbol{\beta}} + \mathbf{D}\boldsymbol{\pi}^{[r]}$. Left multiplying by $\mathbf{D}$ yields
\begin{equation*}
	\mathbf{D}\boldsymbol{\pi}^{[r + 1]} = \PX\tilde{\mathbf{w}}^{[r]} = \tilde{\mathbf{w}}^{[r]} - \MX\tilde{\mathbf{w}}^{[r]}
\end{equation*}
and reveals that we only need to demean $\tilde{\mathbf{w}}^{[r]}$ to update the linear index. 

Finally, we give a short impression about the capabilities of the algorithms presented in this section. Therefore, we estimate the specification from our empirical illustration using a standard routine and the alternative algorithm we suggest. The former requires more than twelve hours whereas our suggestion only needs three seconds.

\section{Simulation Experiments}
\label{sec:sim}

In this section, we extend the analysis of \textcite{fw2016} by two aspects. First, we compare the various analytical and split-panel jackknife bias corrections discussed earlier in this article with respect to their finite sample properties in balanced panels. Second, we analyse whether the improved inference of bias corrections, which is well studied in balanced panels, also shows up in unbalanced panels. For brevity, we restrict ourselves to the analysis of dynamic designs, because \textcite{fw2016} have already shown that the results with respect to exogenous regressors are very similar for static and dynamic designs in balanced panels.

Besides the asymptotically biased maximum likelihood estimator (\textit{MLE}), we consider four different analytical bias corrections for the structural parameters. Two of them are applied at the level of the estimator, whereas the others are the solution of modified score equations. \textit{ABC1} is the analytical bias correction analysed in \textcite{fw2016}. \textit{ABC2} is essentially \textit{ABC1}, but additionally iterated until convergence. \textit{ABC3} and \textit{ABC4} are estimators at the level of the score, whereas the latter recomputes the asymptotic bias in each iteration of the non-linear solver. The analytical bias-corrected estimators of the average partial effects are labelled analogously. Further we analyse the two different splitting strategies (\textit{SPJ1--2}) for the split-panel jackknife bias correction and an alternative estimator for the average partial effects (\textit{LPM}) based on the bias-corrected ordinary least squares estimator proposed by \textcite{hk2002} and the truncated spectral density estimator of \textcite{hk2007}.\footnote{\textcite{hm2006} show that the bias correction of \textcite{hk2002} can be applied to dynamic linear models with individual and time specific effects.} 

\begin{figure}[!htbp]
	\centering
	\caption{\label{fig:missing}Patterns of Randomly Missing Observations.}
	{\includegraphics[width=0.75\textwidth]{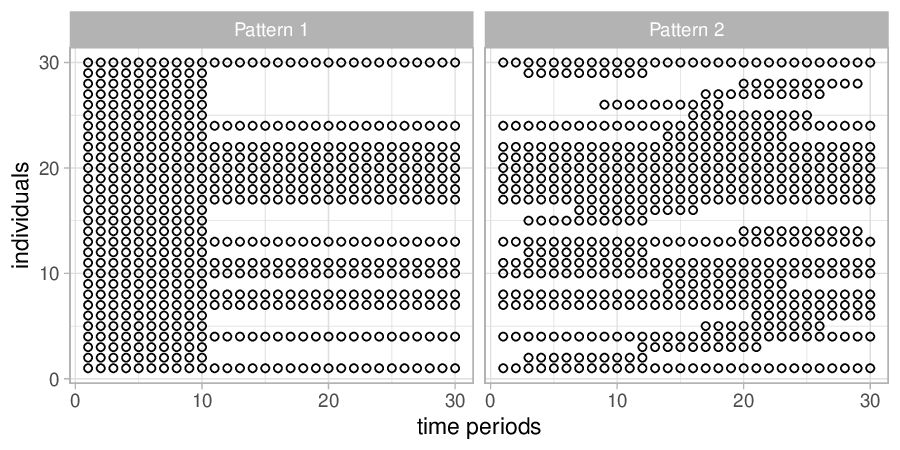} }
\end{figure}

We adapt the dynamic design of \textcite{fw2016} to allow for the possibility of missing data: 
\begin{eqnarray}
	y_{it} &=& \ind\left[\rho y_{it-1} + \beta x_{it} + \alpha_{i} + \gamma_{t} \geq \epsilon_{it} \right] \, , \nonumber \\
	y_{i0} &=& \ind\left[\beta x_{i0} + \alpha_{i} + \gamma_{0} \geq \epsilon_{i0} \right] \, , \nonumber \\
	x_{it} &=& 0.5 x_{it-1} + \alpha_{i} + \gamma_{t} + \nu_{it} \, , \nonumber \\
	(i &=& 1, \ldots, N, \, t = 1 \leq s_{i}, \ldots, T_{i} \leq T) \, , \nonumber
\end{eqnarray}
where $\ind[\cdot]$ is an indicator function, $\alpha_{i}, \gamma_{t} \sim \iid \N(0, 1 / 16)$, $\epsilon_{it} \sim \iid \N(0, 1)$, $\nu_{it} \sim \iid \N(0, 0.5)$, and $x_{i0} \sim \iid \N(0, 1)$. The corresponding parameters are $\rho = 0.5$ and $\beta = 1$. We consider balanced and unbalanced panels with sample sizes that reflect commonly used panel data sets ($N \gg T$). More specifically, we consider two different patterns of randomly missing data that mimic common situations where some people drop out of a survey and are replaced if necessary. To describe the different patterns of missing data, we distinguish between two types of individuals: \textit{type 1} and \textit{type 2}. The former drop out, whereas the latter are observed over the entire time horizon. To be more precise, let $N_{1}$ and $N_{2}$ be the number of \textit{type 1} and \textit{type 2} individuals in the unbalanced panel such that $N = N_{1} + N_{2}$. Likewise $T_{1}$ and $T_{2}$ denote the number of consecutive points in time such that $T_{1} < T_{2}$. Both patterns of missing data differ only in the starting point ($s_{i}$) of the time series of each \textit{type 1} individual. We set $s_{i} = 1$ in \textit{Pattern 1} and sample $s_{i}$ with equal probability from $\{1, \ldots, T_{2} - T_{1} + 1\}$ in \textit{Pattern 2}. For clarification, we set $s_{i} = 1$ for all \textit{type 2} individuals irrespective of the pattern. Figure \ref{fig:missing} provides a graphical illustration of both patterns.
We generate balanced panel data sets with $N = 200$ and $T_{i} = T \in \{15, 20, 25\}$ and unbalanced panel data sets with $(N_{1}, N_{2}) \in \{(300, 100), (150, 150), (60, 180)\}$, $T_{1} = 10$, and $T_{2} = 30$. The different pairs $(N_{1}, N_{2})$ are chosen such that the average number of individuals ($\overline{N}$) and points in time ($\overline{T}$) are $\overline{N} = 200$ and $\overline{T} \in \{15, 20, 25\}$.

To analyse the finite sample performance of the different estimators, we focus on biases relative to the truth and empirical coverage probabilities of 95 \% confidence intervals. The latter statistic is especially important, because even if the relative bias is quite small, it might still be large compared to the dispersion of the estimator, with severe consequences for inference. The \textit{MLE}, \textit{ABC1--4}, and \textit{SPJ1--2} standard errors of the average partial effects are computed using the expression in footnote \ref{fn:vcov}, which takes the independent sampling of the unobserved effects into account. The \textit{LPM} standard errors are based on the cluster-robust covariance estimator of \textcite{cgm2011} to deal with within-individual and within-time correlation of the error terms induced by the probit data generating process. Because there is no obvious way to choose an optimal bandwidth for the estimation of the spectral expectations, we try different choices from a set of values and then report only the results of the choice with the best finite sample performance. We choose $L$ from $\{1, \ldots, 4\}$ for \textit{ABC1--4}, as suggested by \textcite{fw2016, fw2018a}, and from $\{1, \ldots, \overline{T} - 1\}$ for \textit{LPM}. All results are based on 1,000 replications.\footnote{We use the \textit{lfe} package of \textcite{g2013a} for the estimation of linear fixed effects models, the non-linear equations solver (\textit{nleqslv}) of \textcite{h2018} for analytical bias corrections at the level of the scores, and \texttt{R} version 4.0.2 \cite{r2020}.}

We start by comparing the different analytical bias corrections in a balanced panel. Table \ref{tab:anabw} reports the relative biases and coverage probabilities of the estimators for the structural parameters. 
\begin{table}[!htbp]
	\centering
	\begin{threeparttable}
		\caption{\label{tab:anabw}Analytical Bias Corrections: Coefficients.}
		\begin{tabular}{@{}lcccccccc@{}}
			\toprule
			&\multicolumn{4}{c}{$\hat{\rho}$}&\multicolumn{4}{c}{$\hat{\beta}$}\\
			\cmidrule(lr){2-5}\cmidrule(lr){6-9}
			&ABC1&ABC2&ABC3&ABC4&ABC1&ABC2&ABC3&ABC4\\
			\midrule
			\multicolumn{9}{c}{\textit{Panel A: Relative Bias}}\\
			$T = 15$ & -3.886 & -6.580 & -4.673 & -5.204 &  1.004 &  2.666 &  1.710 &  1.761 \\ 
			$T = 20$ & -2.575 & -4.227 & -2.979 & -3.426 &  0.606 &  1.575 &  0.979 &  1.035 \\ 
			$T = 25$ & -1.815 & -2.945 & -2.078 & -2.432 &  0.319 &  0.969 &  0.562 &  0.614 \\ 
			\multicolumn{9}{c}{\textit{Panel B: Coverage Probability}}\\
			$T = 15$ &  0.932 &  0.916 &  0.929 &  0.930 &  0.943 &  0.919 &  0.939 &  0.941 \\ 
			$T = 20$ &  0.947 &  0.931 &  0.944 &  0.939 &  0.941 &  0.930 &  0.938 &  0.941 \\ 
			$T = 25$ &  0.945 &  0.942 &  0.944 &  0.941 &  0.943 &  0.935 &  0.941 &  0.941 \\ 
			\bottomrule
		\end{tabular}
		\begin{tablenotes}
			\footnotesize
			\item\emph{Note:} The biases are in percentage of the truth; \textit{ABC1}--\textit{ABC4} refer to the analytically bias-corrected estimators with bandwidth $L = 2$; results based on 1,000 repetitions.
		\end{tablenotes}
	\end{threeparttable}
\end{table}
As expected, all corrections reduce a larger fraction of the bias and improve coverage as $T$ increases. The difference between the various estimators is in most cases negligible small. Interestingly, \textit{ABC2} does not perform better than \textit{ABC1}, which is remarkable, because often, as for instance in \textcite{ah2007} and \textcite{fw2018a}, it is noted, that a further iteration of \textit{ABC1} could improve the finite sample performance of the estimator.\footnote{\textcite{j2015} analysed an iterated analytical bias correction, similar to \textit{ABC2}, for a static design with only individual unobserved effects with similar findings.} \textit{ABC3} and \textit{ABC4} perform equally well. Our results for the average partial effects are very similar and provided in Table \ref{tab:ana_apes} of the Appendix.

Next we compare the two different split-panel jackknife estimators for the structural parameters in a balanced panel. Table \ref{tab:spj} reports the relative biases and coverage probabilities.
\begin{table}[!htbp]
	\centering
	\begin{threeparttable}
		\caption{\label{tab:spj}Split-Panel Jackknife Bias Corrections: Coefficients.}
		\begin{tabular}{@{}lcccccccc@{}}
			\toprule
			&\multicolumn{4}{c}{Relative Bias}&\multicolumn{4}{c}{Coverage Rate}\\
			\cmidrule(lr){2-5}\cmidrule(lr){6-9}
			&\multicolumn{2}{c}{$\hat{\rho}$}&\multicolumn{2}{c}{$\hat{\beta}$}&\multicolumn{2}{c}{$\hat{\rho}$}&\multicolumn{2}{c}{$\hat{\beta}$}\\
			\cmidrule(lr){2-3}\cmidrule(lr){4-5}\cmidrule(lr){6-7}\cmidrule(lr){8-9}
			&SPJ1&SPJ2&SPJ1&SPJ2&SPJ1&SPJ2&SPJ1&SPJ2\\
			\midrule
			$T = 15$ & -0.030 &  0.540 & -0.744 & -1.620 &  0.907 &  0.907 &  0.895 &  0.884 \\ 
			$T = 20$ &  3.171 &  3.564 & -1.897 & -2.544 &  0.916 &  0.917 &  0.900 &  0.875 \\ 
			$T = 25$ & -0.379 & -0.158 & -0.600 & -0.986 &  0.930 &  0.931 &  0.904 &  0.900 \\ 
			\bottomrule
		\end{tabular}
		\begin{tablenotes}
			\footnotesize
			\item\emph{Note:} The biases are in percentage of the truth; \textit{SPJ1--2} refer to the different split-panel jackknife bias-corrected estimators; results based on 1,000 repetitions. 
		\end{tablenotes}
	\end{threeparttable}
\end{table}
Similar to the analytical bias corrections, we find improved finite sample properties as $T$ increases. One exception is the bias of the estimators in $T = 20$. However, although the bias increases, coverage does not decrease. Overall, we find almost identical properties of both estimators, which is remarkable, because we would expect \textit{SPJ2} to have higher dispersion, as significantly smaller sub panels are used to estimate the asymptotic bias. Results for the average partial effects are similar and reported in Table \ref{tab:spj_apes} of the Appendix.

Now we analyse whether the finite sample properties of the different estimators are affected by the two patterns of randomly missing data. Because we already found that the finite sample properties of the different analytical and split-panel jackknife bias corrections barely differ among themselves, we restrict our final analysis to \textit{MLE}, \textit{ABC1}, \textit{SPJ1}, and \textit{LPM}. Table \ref{tab:propy} and \ref{tab:propx} report the relative biases and coverage probabilities of the estimators for the structural parameters and average partial effects in balanced and unbalanced panels.
\begin{table}[!htbp]
	\centering
	\begin{threeparttable}
		\caption{\label{tab:propy}Finite Sample Properties: Lagged Dependent Variable.}
		\begin{tabular}{@{}llcccccc@{}}
			\toprule
			&&\multicolumn{3}{c}{Relative Bias}&\multicolumn{3}{c}{Coverage Rate}\\
			\cmidrule(lr){3-5}\cmidrule(lr){6-8}
			&&Bal&P1&P2&Bal&P1&P2\\
			\midrule
			\multicolumn{8}{c}{\textit{Panel A: Coefficient}}\\
			$T = \overline{T} = 15$ & MLE & -41.872 & -40.832 & -40.192 & 0.142 & 0.009 & 0.014 \\ 
			& ABC1 & -3.886 & -5.248 & -4.787 & 0.932 & 0.911 & 0.936 \\ 
			& SPJ1 & -0.030 & -32.175 & -19.733 & 0.907 & 0.115 & 0.441 \\ 
			$T = \overline{T} = 20$ & MLE & -31.360 & -30.098 & -29.931 & 0.232 & 0.104 & 0.111 \\ 
			& ABC1 & -2.575 & -2.998 & -2.894 & 0.947 & 0.951 & 0.940 \\ 
			& SPJ1 & 3.171 & -13.741 & -8.846 & 0.916 & 0.668 & 0.822 \\ 
			$T = \overline{T} = 25$ & MLE & -24.997 & -24.405 & -23.860 & 0.306 & 0.263 & 0.278 \\ 
			& ABC1 & -1.815 & -2.057 & -1.554 & 0.945 & 0.949 & 0.942 \\ 
			& SPJ1 & -0.379 & -4.595 & -2.377 & 0.930 & 0.912 & 0.925 \\ 
			\multicolumn{8}{c}{\textit{Panel B: Average Partial Effect}}\\
			$T = \overline{T} = 15$ & MLE & -49.162 & -48.230 & -47.611 & 0.030 & 0.001 & 0.002 \\ 
			& ABC1 (2) & -5.212 & -6.960 & -6.419 & 0.910 & 0.873 & 0.894 \\ 
			& SPJ1 & -10.657 & -39.059 & -26.846 & 0.825 & 0.022 & 0.183 \\ 
			& LPM (4) & 5.269 & 1.800 & 2.153 & 0.913 & 0.961 & 0.930 \\ 
			$T = \overline{T} = 20$ & MLE & -37.768 & -36.550 & -36.496 & 0.083 & 0.017 & 0.018 \\ 
			& ABC1 (2) & -3.200 & -3.816 & -3.849 & 0.911 & 0.918 & 0.910 \\ 
			& SPJ1 & -2.929 & -18.853 & -13.854 & 0.903 & 0.488 & 0.666 \\ 
			& LPM (7) & 4.398 & -0.209 & -0.087 & 0.910 & 0.954 & 0.936 \\ 
			$T = \overline{T} = 25$ & MLE & -30.634 & -30.027 & -29.551 & 0.147 & 0.105 & 0.110 \\ 
			& ABC1 (2) & -2.156 & -2.479 & -2.014 & 0.924 & 0.930 & 0.927 \\ 
			& SPJ1 & -3.944 & -8.102 & -5.813 & 0.892 & 0.832 & 0.873 \\ 
			& LPM (13) & 1.620 & 0.676 & 1.213 & 0.922 & 0.937 & 0.924 \\ 
			\bottomrule
		\end{tabular}
		\begin{tablenotes}
			\footnotesize
			\item\emph{Note:} The biases are in percentage of the truth; \textit{Bal}, \textit{P1}, and \textit{P2} refer to balanced panel, \textit{Pattern 1}, and \textit{Pattern 2}; \textit{MLE}, \textit{ABC1}, \textit{SPJ1}, and \textit{LPM} denote the (bias-corrected) estimators; values in parentheses indicate ``optimal''  bandwidth choices for the spectral density estimator; results based on 1,000 repetitions. 
		\end{tablenotes}
	\end{threeparttable}
\end{table}
\begin{table}[!htbp]
	\centering
	\begin{threeparttable}
		\caption{\label{tab:propx}Finite Sample Properties: Exogenous Regressor.}
		\begin{tabular}{@{}llcccccc@{}}
			\toprule
			&&\multicolumn{3}{c}{Relative Bias}&\multicolumn{3}{c}{Coverage Rate}\\
			\cmidrule(lr){3-5}\cmidrule(lr){6-8}
			&&Bal&P1&P2&Bal&P1&P2\\
			\midrule
			\multicolumn{8}{c}{\textit{Panel A: Coefficient}}\\
			$T = \overline{T} = 15$ & MLE & 14.551 & 13.635 & 13.508 & 0.226 & 0.051 & 0.034 \\ 
			& ABC1 & 1.004 & 0.849 & 0.761 & 0.943 & 0.944 & 0.952 \\ 
			& SPJ1 & -0.744 & 8.995 & 5.790 & 0.895 & 0.343 & 0.640 \\ 
			$T = \overline{T} = 20$ & MLE & 10.657 & 9.838 & 9.962 & 0.329 & 0.205 & 0.205 \\ 
			& ABC1 & 0.606 & 0.301 & 0.400 & 0.941 & 0.943 & 0.948 \\ 
			& SPJ1 & -1.897 & 3.292 & 2.130 & 0.900 & 0.814 & 0.886 \\ 
			$T = \overline{T} = 25$ & MLE & 8.401 & 8.126 & 8.113 & 0.408 & 0.346 & 0.357 \\ 
			& ABC1 & 0.319 & 0.277 & 0.261 & 0.943 & 0.951 & 0.955 \\ 
			& SPJ1 & -0.600 & 0.621 & 0.109 & 0.904 & 0.934 & 0.930 \\ 
			\multicolumn{8}{c}{\textit{Panel B: Average Partial Effect}}\\
			$T = \overline{T} = 15$ & MLE & 2.845 & 2.176 & 2.134 & 0.878 & 0.882 & 0.862 \\ 
			& ABC1 (2) & -0.096 & -0.433 & -0.458 & 0.943 & 0.932 & 0.932 \\ 
			& SPJ1 & 2.617 & 1.744 & 2.797 & 0.870 & 0.819 & 0.812 \\ 
			& LPM (4) & 0.206 & 0.162 & 0.237 & 0.915 & 0.914 & 0.908 \\ 
			$T = \overline{T} = 20$ & MLE & 2.285 & 1.731 & 1.693 & 0.881 & 0.874 & 0.890 \\ 
			& ABC1 (2) & 0.044 & -0.271 & -0.316 & 0.942 & 0.924 & 0.925 \\ 
			& SPJ1 & 1.364 & 0.946 & 1.494 & 0.909 & 0.896 & 0.895 \\ 
			& LPM (7) & 0.166 & 0.312 & 0.225 & 0.923 & 0.912 & 0.924 \\ 
			$T = \overline{T} = 25$ & MLE & 1.848 & 1.675 & 1.614 & 0.888 & 0.903 & 0.890 \\ 
			& ABC1 (2) & 0.015 & -0.048 & -0.112 & 0.932 & 0.941 & 0.933 \\ 
			& SPJ1 & 0.889 & 0.735 & 0.836 & 0.917 & 0.924 & 0.908 \\ 
			& LPM (13) & 0.402 & 0.395 & 0.276 & 0.916 & 0.927 & 0.912 \\ 
			\bottomrule
		\end{tabular}
		\begin{tablenotes}
			\footnotesize
			\item\emph{Note:} The biases are in percentage of the truth; \textit{Bal}, \textit{P1}, and \textit{P2} refer to balanced panel, \textit{Pattern 1}, and \textit{Pattern 2}; \textit{MLE}, \textit{ABC1}, \textit{SPJ1}, and \textit{LPM} denote the (bias-corrected) estimators; values in parentheses indicate ``optimal'' bandwidth choices for the spectral density estimator; results based on 1,000 repetitions.
		\end{tablenotes}
	\end{threeparttable}
\end{table}
We start with the analysis of the finite sample properties in a balanced panel, as these will serve as a benchmark for the properties in unbalanced panels. First, we find that the estimators for the effects of $y_{it - 1}$ are worse than those for $x_{it}$. For instance, we observe larger biases and coverage probabilities that are further away from their nominal level. The distortions in the coefficients are also apparent in the estimates of the average partial effects, which is in contrast to the negligible small biases in the average partial effects of $x_{it}$.\footnote{\textcite{hn2004}, \textcite{f2009}, and \textcite{fw2016} have similar findings for average partial effects of an exogenous regressor in balanced panels.} In general, the bias corrections perform well in reducing the relative biases and improving coverage. As in \textcite{fw2016}, the properties of \textit{SPJ1} are worse than those of \textit{ABC1}. \textit{LPM} as an alternative estimator for the average partial effects works well too.\footnote{\textcite{f2009} has similar findings for linear probability models with only individual fixed effects.} Interestingly, \textit{LPM} tends to overestimate the average partial effects of $y_{it - 1}$, while the other estimators underestimate them.	Further, the optimal bandwidths of \textit{LPM} for the estimation of the spectral expectations are much larger than for \textit{ABC1} and increase rapidly with $T$. This indicates that the temporal dependence induced by the probit data generating process can be very strong, which in turn makes the choice of appropriate bandwidths for \textit{LPM} difficult in practice. Next we compare the finite sample properties of the different estimators in unbalanced panels with our benchmark. First, by comparing the relative biases of \textit{MLE}, we can confirm the conjecture of \textcite{fw2018a} that the magnitude of the asymptotic bias in the limiting distribution depends on $\overline{N}$ and $\overline{T}$. However, the coverage is worse because the sample size of an unbalanced panel is larger than that of a balanced panel (with equal $\overline{N}$ and $\overline{T}$), which in turn implies a smaller standard deviation of the estimators and thus distortions that are larger relative to the variance of the estimator. Second, compared to the benchmark we notice some substantial differences in the performance of \textit{SPJ1}. Whereas the properties of \textit{ABC1} and \textit{LPM} are unaffected by the different patterns of missing data, they are partially significantly worse for \textit{SPJ1}, especially for $T = \overline{T} = 15$. \textit{Pattern 1} stands out in particular, because it clearly shows that the reduction of bias and improvement of coverage are deteriorating. An intuitive explanation is that the splitting strategy leads to sub panels of widely differing sizes. This issue is not so severe in \textit{Pattern 2}, but the performance is still worse than the benchmark.

Finally, we briefly summarize the key findings. We find no differences in the finite sample performance when we compare the various analytically bias-corrected and the different split-panel jackknife estimators among themselves. Although the latter have the advantage that they are relatively easy to implement, this convenience is associated with some performance losses. More precisely, split-panel jackknife estimators have higher distortion and react sensitive to different patterns of randomly missing data, whereas the performance of the analytical bias corrections is unaffected. An alternative estimator for the average partial effects based on the bias-corrected ordinary least squares estimator works well too, but to find an appropriate bandwidth for the required spectral density estimator might be challenging in practice.

In the next section, we demonstrate the usefulness of computational advances and bias corrections with an empirical example from labour economics.

\section{Empirical Illustration}
\label{sec:illustration}

We illustrate one possible area of application by analysing the inter-temporal labour force participation of women using longitudinal micro data (1984--2013) from the German Socio Economic Panel (\textit{GSOEP}). More specifically, we want to investigate how fertility decisions and non-labour income jointly affect women's labour force participation using an unbalanced panel data set of 6,241 women in labour force observed consecutively for at least ten years. Further details about the sample are provided in the Appendix.

In spirit of \textcite{h1999}, we estimate the following model specification: 
\begin{eqnarray}
	y_{it} &=& \mathbf{1}\left[\rho y_{it - 1} + \mathbf{x}_{it}^{\prime} \boldsymbol{\beta} + \alpha_{i} + \gamma_{t} + e_{it} \geq 0 \right] \, , \nonumber \\
	(i &=& 1, \ldots, N, \, t = 1 \leq s_{i}, \ldots, T_{i} \leq T) \, , \nonumber
\end{eqnarray}
where $y_{it}$ is an indicator equal to one if woman $i$ participates in the labour force at time $t$, $\mathbf{x}_{it}$ is a vector of explanatory and further control variables,  $\boldsymbol{\beta}$ are the corresponding common parameters, $\alpha_{i}$ and $\gamma_{t}$ are unobserved effects that capture individual specific taste for labour and permanent income as well as control for the business cycle and other time specific shifts in preferences, and $e_{it}$ is an idiosyncratic error term independent of $\mathbf{X}_{i}^{\star}$ and $\boldsymbol{\pi}$ with mean zero. As our data set is unbalanced, we have to additionally assume that conditional on $\mathbf{X}_{i}^{\star}$ and $\boldsymbol{\pi}$ the attrition process is random. We consider the following explanatory variables: number of children in different age groups, various non-labour income classes, and an indicator that is equal to one if a birth occurs in the next year. Further control variables are squared age, marital status, a regional identifier for Eastern Germany, number of children between zero and one in the previous year, and number of other household members.

Table \ref{tab:summary} shows some descriptive statistics of our sample.
\begin{table}[!htbp]
	\centering
	\begin{threeparttable}
		\caption{\label{tab:summary}Descriptive Statistics.}
		\begin{tabular}{@{}lcccc@{}}
			\toprule
			&Full&Always&Never&Movers\\
			\midrule
			Participation & 0.73 (0.44) & 1.00 (0.00) & 0.00 (0.00) & 0.67 (0.47) \\ 
			Age & 40.48 (11.02) & 42.69 (9.80) & 48.02 (10.65) & 38.21 (11.06) \\ 
			Married & 0.70 (0.46) & 0.67 (0.47) & 0.93 (0.26) & 0.69 (0.46) \\ 
			Middle Class & 0.44 (0.50) & 0.42 (0.49) & 0.48 (0.50) & 0.44 (0.50) \\ 
			Upper Class & 0.01 (0.08) & 0.01 (0.08) & 0.01 (0.10) & 0.01 (0.08) \\ 
			East & 0.22 (0.41) & 0.28 (0.45) & 0.05 (0.22) & 0.20 (0.40) \\ 
			\#Children 0-1 & 0.05 (0.21) & 0.02 (0.13) & 0.04 (0.21) & 0.06 (0.25) \\ 
			\#Children 2-4 & 0.12 (0.36) & 0.05 (0.23) & 0.12 (0.36) & 0.16 (0.41) \\ 
			\#Children 5-18 & 0.71 (0.95) & 0.53 (0.80) & 0.81 (1.14) & 0.79 (0.99) \\ 
			\#HH older & 2.25 (0.84) & 2.18 (0.80) & 2.61 (0.94) & 2.25 (0.84) \\ 
			Birth$_{t + 1}$ & 0.03 (0.18) & 0.01 (0.11) & 0.03 (0.16) & 0.04 (0.21) \\ 
			\midrule
			\#Observations&97,465&33,574& 7,175&56,716\\
			\#Individuals ($N$)&6,241&2,306&  477&3,458\\
			\#Years ($T$)&28&28&28&28\\
			Avg. \#Individuals ($\overline{N}$)&3,481&1,199&  256&2,026\\
			Avg. \#Years ($\overline{T}$)&16&15&15&16\\
			\bottomrule
		\end{tabular}
		\begin{tablenotes}
			\footnotesize
			\item\emph{Source:} \textit{GSOEP} 1984--2013.
		\end{tablenotes}
	\end{threeparttable}
\end{table}
The average participation rate is 73 \% in the full sample and 67 \% for the group of movers who change their labour force participation decision at least once. The group of women who never participate is the smallest and most different from the other groups. On average, this group is older, more likely to be married, and lives in Western Germany. Contrary, women who always participate have less children and live in smaller households. The full sample comprises 97,465 observations and consists of 6,241 women observed for a maximum of 28 years. As some households drop out of the \textit{GSOEP} and are replaced by new ones, this leads to a pattern of missing data similar to \textit{Pattern 2} in the simulation study. On average, we observe each woman for roughly 16 years, or each year we observe about 3,481 women.

We consider the following estimators  for the structural parameters and average partial effects: \textit{MLE}, \textit{ABC1}, \textit{SPJ1}, and \textit{LPM}. The estimators are labelled and bandwidths are chosen as in the simulation study. Table \ref{tab:lfp} reports the corresponding estimates.
\begin{table}[!htbp]
	\centering
	\begin{threeparttable}
		\caption{\label{tab:lfp}Labor Force Participation of Women.}
		\begin{tabular}{@{}lccccccc@{}}
			\toprule
			&\multicolumn{3}{c}{Coefficient}&\multicolumn{4}{c}{Average Partial Effect}\\
			\cmidrule(lr){2-4}\cmidrule(lr){5-8}
			&MLE&ABC1&SPJ1&MLE&ABC1&SPJ1&LPM\\
			\midrule
			Participation$_{t - 1}$ &  1.469 &  1.657 &  1.712 &  0.233 &  0.300 &  0.307 &  0.540 \\ 
			& (0.017) & (0.017) & (0.017) & (0.019) & (0.023) & (0.006) & (0.015) \\ 
			Middle Class & -0.122 & -0.112 & -0.114 & -0.013 & -0.013 & -0.014 & -0.014 \\ 
			& (0.023) & (0.023) & (0.023) & (0.003) & (0.003) & (0.002) & (0.003) \\ 
			Upper Class & -0.365 & -0.330 & -0.416 & -0.042 & -0.041 & -0.051 & -0.034 \\ 
			& (0.130) & (0.132) & (0.132) & (0.017) & (0.016) & (0.016) & (0.016) \\ 
			\#Children 0-1 & -1.822 & -1.603 & -1.704 & -0.198 & -0.190 & -0.206 & -0.304 \\ 
			& (0.036) & (0.035) & (0.036) & (0.017) & (0.015) & (0.006) & (0.010) \\ 
			\#Children 2-4 & -0.427 & -0.304 & -0.335 & -0.046 & -0.036 & -0.042 & -0.039 \\ 
			& (0.026) & (0.025) & (0.026) & (0.005) & (0.004) & (0.003) & (0.005) \\ 
			\#Children 5-18 & -0.159 & -0.113 & -0.112 & -0.017 & -0.013 & -0.014 & -0.014 \\ 
			& (0.012) & (0.012) & (0.013) & (0.002) & (0.002) & (0.001) & (0.002) \\ 
			Birth$_{t + 1}$ & -0.560 & -0.514 & -0.572 & -0.065 & -0.066 & -0.074 & -0.080 \\ 
			& (0.038) & (0.038) & (0.038) & (0.007) & (0.007) & (0.005) & (0.007) \\ 
			\bottomrule
		\end{tabular}
		\begin{tablenotes}
			\footnotesize
			\item\emph{Note:} \textit{MLE}, \textit{ABC1}, \textit{SPJ1}, and \textit{LPM} denote the (bias-corrected) estimators; bandwidths are 2 and 4 for \textit{ABC1} and \textit{LPM}; standard errors in parentheses; \textit{LPM} standard errors are robust to heteroskedasticity and clustered by woman and year; estimates relative to a low income woman in the west.
			\item\emph{Further control variables:} squared age, married, east, lag of number of children between zero and one, and number of household members above 18.
			\item\emph{Source:} \textit{GSOEP} 1984--2013.
		\end{tablenotes}
	\end{threeparttable}
\end{table}
All results are intuitive and in line with the theoretical model of \textcite{h1999}. We find positive state dependence and negative effects of transitory non-labour income, number of children, and expectations about future fertility. All effects are significant at the 5 \% level. Remarkably, most probit estimates of the average partial effects are very close to each other. Exceptions are those with respect to lagged participation which range from 0.23 up to 0.31. \textit{LPM} estimates are also quite close except for lagged participation and number of children between zero and one. Overall we find evidence for strong state dependence. A woman who has currently a job increases her probability to participate in the future by 23--54 percentage points. Further we find that women respond heterogeneously to changes in transitory non-labour income. Being in the middle class reduces the participation probability by roughly one percentage point compared to a woman in the lower income class. The reduction associated with belonging to the upper income class is significantly stronger with three up to five percentage points. Finally we find that the number of children reduces the likelihood of participation substantially. As expected, the effect is declining in age of children. Each additional child between zero and one reduces the probability to participation 20 up to 30 percentage points. For children older than four, the reduction is only one percentage point. The results  are largely consistent with the empirical findings of \textcite{h1999}. However, contrary to him, we find that future birth always negatively affects current participation decision irrespective of the chosen estimator. This might support the author's perfect foresight assumption with respect to life-cycle fertility decisions.

Finally, we check the sensitivity of \textit{ABC1} and \textit{LPM} to different bandwidth choices and conduct a simulation study calibrated to our empirical illustration. The results are reported in Table \ref{tab:sensitivity} and \ref{tab:calibrated} of the Appendix. With respect to the different bandwidth choices, we find that especially the \textit{ABC1} estimates are very robust. We find the largest variation with respect to the effect of lagged participation. The other effects are almost indistinguishable. Most \textit{LPM} estimates are robust as well. Exceptions are the effects of lagged participation, being in the upper class, and number of children between zero and one. The results of the calibrated simulation study confirm that the finite sample properties of \textit{MLE} can be improved. However, the improvement is less good compared to the simulation experiments with synthetic data. The performance of \textit{SPJ1} and \textit{LPM} is in some cases significantly worse than that of \textit{ABC1}. Given that there have been several labour market reforms with heterogeneous effects on different sub populations that most likely violate the unconditional homogeneity assumption of \textit{SPJ1}, this could partly explain its worse performance.

\section{Concluding Remarks}
\label{sec:conclusion}

In this article, we offer new relief and guidance for empirical researchers by showing how popular binary choice estimators benefit from recent advances in econometrics. Especially the analytically bias-corrected estimator of \textcite{fw2016} convinced by its good performance in all cases.

Although we have focused on panel data binary choice models, we would like to point out that the bias corrections derived by \textcite{fw2016} are far more general. First, they can also be used to reduce the asymptotic bias of other popular non-linear maximum likelihood estimators e. g. poisson and tobit. The algorithms described in this article can be easily adapted to these problems. Second, the bias corrections can also be applied if we observe cross-sections of networked activities instead of panels e. g. a cross-section of bilateral trade flows. For instance, \textcite{cfw2017} use the bias corrections of \textcite{fw2016} to mitigate the asymptotic bias problem in a \textcite{hmr2008}-type model to determine the extensive margin of trade.

Recently, there are also some extensions of \textcite{fw2016}. For instance, \textcite{wz2020} and \textcite{hsw2020} have extended these bias corrections to a special three-way error component that is particularly relevant for the estimation of the intensive and extensive margin of trade in a panel of bilateral trade flows. Further, \textcite{cfw2020} use multiple binary choice regressions to estimate the distribution of non-binary outcomes conditional on strictly exogenous regressors and two unobserved effects. The corresponding inference problem is addressed using an analytical bias correction. These extensions can also benefit from the findings of this article.

Finally, future research could investigate whether bootstrap procedures can further improve inference. In an additional simulation study we find that bootstrapping a bias-corrected estimator in spirit of \textcite{k2014} slightly improves the coverage of the estimators for the structural parameters but not for the average partial effects, which questions the validity of this bootstrap procedure in our specific setting.

\printbibliography

\clearpage

\appendix

{\LARGE\textbf{Appendix}}
\captionsetup{labelformat=AppendixTables}
\setcounter{table}{0}

\section{Feasible Estimation for Long Panels}

\subsection{Derivation of the Algorithms}

We briefly review the derivation of the algorithm proposed by \textcite{s2018} for binary choice models with two unobserved effects. 

We start with the weighted least squares problem in iteration $r$:
\begin{equation}
	\label{eq:iwls}
	\boldsymbol{\theta}^{[r + 1]} = (\boldsymbol{\beta}^{[r + 1]}, \boldsymbol{\pi}^{[r + 1]}) = (\mathbf{Z}^{\prime} \boldsymbol{\Omega}^{[r]} \mathbf{Z})^{+} \mathbf{Z}^{\prime} \boldsymbol{\Omega}^{[r]} \mathbf{w}^{[r]} \, ,
\end{equation}
where $\boldsymbol{\pi}^{[r]} = (\boldsymbol{\alpha}^{[r]}, \boldsymbol{\gamma}^{[r]})$, $\mathbf{Z} = (\mathbf{X}, \mathbf{D})$, $\boldsymbol{\Omega}^{[r]}$ is a diagonal weighting matrix with $\text{diag}(\boldsymbol{\Omega}^{[r]}) = (\omega_{it}^{[r]})_{(i, t) \in \mathcal{S}}$, $\mathbf{w}^{[r]} = \boldsymbol{\nu}^{[r]} + \boldsymbol{\eta}^{[r]}$, $\boldsymbol{\nu}^{[r]} = (\nu_{it}^{[r]})_{(i, t) \in \mathcal{S}}$, and $\boldsymbol{\eta}^{[r]} = (\eta_{it}^{[r]})_{(i, t) \in \mathcal{S}}$. The corresponding expressions for $\omega_{it}$ and $\nu_{it}$ are given in section 2.2. \eqref{eq:iwls} implies the following normal equations:
\begin{eqnarray}
	\mathbf{X}^{\prime} \boldsymbol{\Omega}^{[r]} \mathbf{X} \boldsymbol{\beta}^{[r + 1]} + \mathbf{X}^{\prime} \boldsymbol{\Omega}^{[r]} \mathbf{D} \boldsymbol{\pi}^{[r + 1]} &=& \mathbf{X}^{\prime} \boldsymbol{\Omega}^{[r]} \mathbf{w}^{[r]} \label{eq:neq1} \, , \\
	\mathbf{D}^{\prime} \boldsymbol{\Omega}^{[r]} \mathbf{X} \boldsymbol{\beta}^{[r + 1]} + \mathbf{D}^{\prime} \boldsymbol{\Omega}^{[r]} \mathbf{D} \boldsymbol{\pi}^{[r + 1]} &=& \mathbf{D}^{\prime} \boldsymbol{\Omega}^{[r]} \mathbf{w}^{[r]} \label{eq:neq2} \, .
\end{eqnarray}
Re-arranging \eqref{eq:neq2} yields
\begin{equation}
	\label{eq:neq3}
	\mathbf{D} \boldsymbol{\pi}^{[r + 1]} = \PX\mathbf{w}^{[r]} - \PX\mathbf{X}^{[r]} \boldsymbol{\beta}^{[r + 1]}
\end{equation}
Substituting \eqref{eq:neq3} into \eqref{eq:neq1} results in an alternative weighted least squares problem
\begin{equation}
	\label{eq:ciwls}
	\boldsymbol{\beta}^{[r + 1]} = ((\MX\mathbf{X}^{[r]})^{\prime} \boldsymbol{\Omega}^{[r]} \MX\mathbf{X}^{[r]})^{- 1} (\MX\mathbf{X}^{[r]})^{\prime} \boldsymbol{\Omega}^{[r]} \MX\mathbf{w}^{[r]} \, .
\end{equation}
based on transformed variables that allows us to update $\boldsymbol{\beta}$ without explicitly updating the incidental parameters $\boldsymbol{\pi}$. This transformation can be interpreted as a weighted within-transformation (``demeaning'') as known from linear fixed effects models. Consequently, $\boldsymbol{\beta}$ can be obtained by a regression of the transformed $\mathbf{w}$ on the transformed $\mathbf{X}$ using $\boldsymbol{\omega}$ as weights. Meanwhile, this type of optimization problem can be solved very efficiently thanks to special software routines. Some examples available in popular statistical software are \textit{reghdfe} by \textcite{c2016} for \textit{Stata}, \textit{lfe} by \textcite{g2013a} for \textit{R}, \textit{FixedEffectModels} by Matthieu Gomez for \textit{Julia}, and \textit{pyhdfe} by Jeff Gortmaker for \textit{Python}. To complete the optimization algorithm we need to find a way to update $\omega_{it}$ and $\nu_{it}$. Fortunately, it is quite easy to update $\eta_{it}$ from already computed quantities. We can exploit the fact that the residuals of \eqref{eq:iwls} and \eqref{eq:ciwls} are equal, as shown by \textcite{g2013b}. Therefore
\begin{equation}
	\label{eq:residuals}
	\boldsymbol{\eta}^{[r + 1]} = \mathbf{w}^{[r]} - (\MX\mathbf{w}^{[r]} - \MX\mathbf{X}^{[r]} \boldsymbol{\beta}^{[r + 1]})
\end{equation}
are simply the fitted values of \eqref{eq:ciwls}. We can sketch the entire algorithm as follows:
\begin{definition}
	IWLS Algorithm
	
	Set an initial $\boldsymbol{\eta}$, e. g. $\boldsymbol{\eta} = \mathbf{0}_{n}$,  and repeat the following steps.  
	\begin{itemize}[itemindent=3em]
		\item[Step 1:] Compute $\boldsymbol{\omega}$ and $\mathbf{w}$.
		\item[Step 2:] Solve \eqref{eq:ciwls}, where $\boldsymbol{\beta}$ and $\boldsymbol{\eta}$ are the coefficients and fitted values.
		\item[Step 3:] Check convergence.
	\end{itemize}
\end{definition}

Finally we outline the algorithm for updating the linear index $\eta_{it}$ given a fixed $\tilde{\boldsymbol{\beta}}$. For instance, this algorithm is required for the bias corrections of the average partial effects. Remember, 
\begin{equation*}
	\mathbf{D}\boldsymbol{\pi}^{[r + 1]} = \PX\tilde{\mathbf{w}}^{[r]} = \tilde{\mathbf{w}}^{[r]} - \MX\tilde{\mathbf{w}}^{[r]}
\end{equation*}
with $\tilde{\mathbf{w}}^{[r]} = \mathbf{w}^{[r]} - \mathbf{X} \tilde{\boldsymbol{\beta}}$ is sufficient to update $\boldsymbol{\eta}^{[r]} = \mathbf{X}\tilde{\boldsymbol{\beta}} + \mathbf{D}\boldsymbol{\pi}^{[r]}$. We can sketch the entire algorithm as follows:
\begin{definition}
	IWLS Offset Algorithm
	
	Set an initial $\boldsymbol{\eta}$, e. g. $\boldsymbol{\eta} = \mathbf{0}_{n}$,  and repeat the following steps.
	\begin{itemize}[itemindent=3em]
		\item[Step 1:] Compute $\boldsymbol{\omega}$ and $\mathbf{w}^{\ast}$.
		\item[Step 2:] Compute $\mathbf{D} \boldsymbol{\pi}$ and update $\boldsymbol{\eta}$.
		\item[Step 3:] Check convergence.
	\end{itemize}
\end{definition}

In the next subsection we present an example code in which we combine the algorithms described here with the analytical bias correction of \textcite{fw2016}.

\clearpage

\subsection{Example Code}

\begin{verbatim}
########################################################
# Example R-Code for Logit MLE
########################################################


## Load required packages and data set
require(bife)
require(lfe)
data <- psid

## Helper functions

# Log-Likelihood function
logl <- function(y, eta) sum(log(plogis((2.0 * y - 1.0) * eta)))

# Derivatives of CDF
partial2F <- function(eta) dlogis(eta) * (1.0 - 2.0 * plogis(eta))
partial3F <- function(eta) {
    dlogis(eta) * ((1.0 - 2.0 * plogis(eta))^2 - 2.0 * dlogis(eta))
}

## Estimate structural parameters

# Set initial \beta, \eta, and termination criteria
n <- nrow(data)
fval <- - 1.0e100
eta <- numeric(n)
maxiter <- 100L
tol <- 1.0e-05

# Find optimal \beta and \eta
for (iter in 1:maxiter) {
    # Store previous function value
    fvalold <- fval

    # Update weights and working response
    omega <- dlogis(eta)
    data$w <- (data$LFP - plogis(eta)) / omega + eta

    # Update \beta and \eta
    reg <- felm(w ~ KID1 + KID2 + KID3 + log(INCH) | ID + TIME,
    data, weights = omega)
    beta <- coef(reg)
    eta <- as.vector(fitted.values(reg))
    fval <- logl(data$LFP, eta)

    # Check convergence
    if (abs(fval - fvalold) / (abs(fvalold) + 0.1) < tol) break
}

# Final estimates
betahat <- beta
etahat <- eta

## Debias maximum likelihood estimates of the structural parameters

# Compute required derivatives and demeaned regressor matrix
d1lhat <- data$LFP - plogis(etahat)
omegahat <- dlogis(etahat)
d2Fhat <- partial2F(etahat)
reg <- felm(KID1 + KID2 + KID3 + log(INCH) ~ 0 | ID + TIME,
data, weights = omegahat)
MhatX <- residuals(reg)

# Estimate asymptotic bias
Tempmat <- MhatX * d2Fhat
Bhatnum <- aggregate(Tempmat, list(data$ID), sum)[, - 1L]
Bhatdenom <- aggregate(omegahat, list(data$ID), sum)[, - 1L]
Bhat <- - colSums(Bhatnum / Bhatdenom) / 2.0
Chatnum <- aggregate(Tempmat, list(data$TIME), sum)[, - 1L]
Chatdenom <- aggregate(omegahat, list(data$TIME), sum)[, - 1L]
Chat <- - colSums(Chatnum / Chatdenom) / 2.0
What <- crossprod(MhatX * sqrt(omegahat))
bhat <- solve(What, Bhat + Chat)

# Estimate asymptotic covariance matrix
Vhat <- solve(What)

# Debias estimates and compute standard errors
betatilde <- betahat - bhat
sebetatilde <- sqrt(diag(Vhat))

## Debias maximum likelihood estimates of the APEs

# Compute fixed part of \eta
X <- model.matrix(LFP ~ KID1 + KID2 + KID3 + log(INCH) + 0, data)
Xbetatilde <- as.vector(X %*% betatilde)

# Set initial \eta and termination criteria
fval <- - 1.0e100
eta <- numeric(n)

# Update \eta given debiased \beta
for (iter in 1:maxiter) {
    # Store previous function value
    fvalold <- fval

    # Update weights and working response
    omega <- dlogis(eta)
    data$wast <- (data$LFP - plogis(eta)) / omega + eta - Xbetatilde

    # Update \eta
    reg <- felm(wast ~ 0 | ID + TIME, data, weights = omega)
    Dpi <- as.vector(fitted.values(reg))
    eta <- Xbetatilde + Dpi
    fval <- logl(data$LFP, eta)

    # Check convergence
    if (abs(fval - fvalold) / (abs(fvalold) + 0.1) < tol) break
}

# Final estimates
etatilde <- eta

# Compute APEs and derivatives evaluated at debiased coefficients
Deltahat <- sapply(betatilde, function(x) x * dlogis(etatilde))
d1Deltahat <- sapply(betatilde, function(x) x * partial2F(etatilde))
d2Deltahat <- sapply(betatilde, function(x) x * partial3F(etatilde))
deltahat <- colMeans(Deltahat)
Psihat <- - d1Deltahat / omegahat
reg <- felm(Psihat ~ 0 | ID + TIME, data, weights = omegahat)
PhatPsihat <- fitted.values(reg)

# Estimate asymptotic bias
Tempmat <- PhatPsihat * d2Fhat + d2Deltahat 
Bhatnum <- aggregate(Tempmat, list(data$ID), sum)[, - 1L]
Bhatdenom <- aggregate(omegahat, list(data$ID), sum)[, - 1L]
Bhat <- colSums(Bhatnum / Bhatdenom) / 2.0
Chatnum <- aggregate(Tempmat, list(data$TIME), sum)[, - 1L]
Chatdenom <- aggregate(omegahat, list(data$TIME), sum)[, - 1L]
Chat <- colSums(Chatnum / Chatdenom) / 2.0
bhat <- (Bhat + Chat) / n

# Estimate asymptotic covariance matrix
V1 <- tcrossprod(colSums(Deltahat - deltahat))
J <- crossprod(MhatX, d1Deltahat)
diag(J) <- diag(J) + sum(dlogis(etatilde))
JV <- t(J) %*% solve(What)
Gammahat <- tcrossprod(MhatX * d1lhat, JV) - PhatPsihat * d1lhat
V2 <- crossprod(Gammahat)
Vhat <- (V1 + V2) / n^2

# Debias estimates and compute standard errors
deltatilde <- deltahat - bhat
sedeltatilde <- sqrt(diag(Vhat))
\end{verbatim}

\clearpage

\section{Additional Simulation Results}

\begin{table}[!htbp]
	\centering
	\begin{threeparttable}
		\caption{\label{tab:ana_apes}Analytical Bias Corrections: Average Partial Effects.}
		\begin{tabular}{@{}lcccccccc@{}}
			\toprule
			&\multicolumn{4}{c}{$\hat{\delta}_{y}$}&\multicolumn{4}{c}{$\hat{\delta}_{x}$}\\
			\cmidrule(lr){2-5}\cmidrule(lr){6-9}
			&ABC1&ABC2&ABC3&ABC4&ABC1&ABC2&ABC3&ABC4\\
			\midrule
			\multicolumn{9}{c}{\textit{Panel A: Relative Bias}}\\
			$T = 15$ & -5.212 & -8.492 & -6.248 & -6.805 & -0.096 &  1.050 &  0.379 &  0.433 \\ 
			$T = 20$ & -3.200 & -5.223 & -3.744 & -4.221 &  0.044 &  0.725 &  0.297 &  0.352 \\ 
			$T = 25$ & -2.156 & -3.543 & -2.512 & -2.892 &  0.015 &  0.477 &  0.181 &  0.230 \\ 
			\multicolumn{9}{c}{\textit{Panel B: Coverage Probability}}\\
			$T = 15$ &  0.910 &  0.874 &  0.897 &  0.895 &  0.943 &  0.930 &  0.941 &  0.943 \\ 
			$T = 20$ &  0.911 &  0.900 &  0.908 &  0.910 &  0.942 &  0.933 &  0.941 &  0.939 \\ 
			$T = 25$ &  0.924 &  0.920 &  0.923 &  0.925 &  0.932 &  0.927 &  0.928 &  0.928 \\ 
			\bottomrule
		\end{tabular}
		\begin{tablenotes}
			\footnotesize
			\item\emph{Note:} The biases are in percentage of the truth; \textit{ABC1}--\textit{ABC4} refer to the analytically bias-corrected estimators with bandwidth $L = 2$; results based on 1,000 repetitions.
		\end{tablenotes}
	\end{threeparttable}
\end{table}

\begin{table}[!htbp]
	\centering
	\begin{threeparttable}
		\caption{\label{tab:spj_apes}Split-Panel Jackknife Bias Corrections: Average Partial Effects.}
		\begin{tabular}{@{}lcccccccc@{}}
			\toprule
			&\multicolumn{4}{c}{Relative Bias}&\multicolumn{4}{c}{Coverage Rate}\\
			\cmidrule(lr){2-5}\cmidrule(lr){6-9}
			&\multicolumn{2}{c}{$\hat{\delta}_{y}$}&\multicolumn{2}{c}{$\hat{\delta}_{x}$}&\multicolumn{2}{c}{$\hat{\delta}_{y}$}&\multicolumn{2}{c}{$\hat{\delta}_{x}$}\\
			\cmidrule(lr){2-3}\cmidrule(lr){4-5}\cmidrule(lr){6-7}\cmidrule(lr){8-9}
			&SPJ1&SPJ2&SPJ1&SPJ2&SPJ1&SPJ2&SPJ1&SPJ2\\
			\midrule
			$T = 15$ & -10.657 & -10.609 &   2.617 &   2.460 &   0.825 &   0.825 &   0.870 &   0.874 \\ 
			$T = 20$ &  -2.929 &  -2.890 &   1.364 &   1.241 &   0.903 &   0.898 &   0.909 &   0.913 \\ 
			$T = 25$ &  -3.944 &  -3.927 &   0.889 &   0.817 &   0.892 &   0.890 &   0.917 &   0.914 \\ 
			\bottomrule
		\end{tabular}
		\begin{tablenotes}
			\footnotesize
			\item\emph{Note:} The biases are in percentage of the truth; \textit{SPJ1--2} refer to the different split-panel jackknife bias-corrected estimators; results based on 1,000 repetitions. 
		\end{tablenotes}
	\end{threeparttable}
\end{table}

\clearpage

\section{Empirical Illustration}

\subsection{Data Preparation}

We use the \textit{Cross-National Equivalent File} (\$PEQUV-File version 30) of the \textit{GSOEP} and restrict the sample to women between 16 and 65 that are observed consecutively for at least ten years and do not receive any retirement income. A woman is assumed to participate in labour-force if she has positive income from individual labour and works at least 52 hours a year. A proxy for transitory non-labour income is constructed from post-government household income minus woman's individual labour earnings. All income variables are converted to constant 2010 EURO using the consumer price index. We additionally correct the labour income by a household specific tax rate. To make income comparable between different household sizes, we use an equivalence scale proposed by \textcite{brss1988} and divide the transitory non-labour income by the square root of household members. To allow for heterogeneous effects of transitory non-labour income on participation decisions, we define the three income classes: lower, middle, and upper. A woman belongs to the lower class if she has a non-labour income of less than 11,278 EURO at her disposal. Contrary a woman is in the upper income class if she has more than 56,391 EURO available. Women in between this interval belong to the middle class. Those numbers are equal to 60 \% and 300 \% of the annual median equivalence income.\footnote{\url{https://www.destatis.de/DE/Themen/Gesellschaft-Umwelt/Einkommen-Konsum-Lebensbedingungen/Lebensbedingungen-Armutsgefaehrdung/Tabellen/einkommensverteilung-silc.html}.} The class distinction is taken from the \textit{Armuts- und Reichtumsbericht} of the federal government.\footnote{\url{https://www.armuts-und-reichtumsbericht.de/}.
} Finally, we group the federal states into Eastern and Western Germany to control for regional differences that still exists after reunification. More precisely, Schleswig-Holstein, Hamburg, Lower-Saxony, Bremen, Hessen, Baden-Wuerttemberg, Bavaria, North-Rhine-Westfalia, Rheinland-Pfalz, and Saarland are classified as Western Germany and Berlin, Brandenburg, Mecklenburg-Vorpommern, Saxony, Saxony-Anhalt, and Thueringia as Eastern Germany.

\clearpage

\subsection{Additional Results}

\begin{table}[!htbp]
	\centering
	\begin{threeparttable}
		\caption{\label{tab:sensitivity}Sensitivity to Different Bandwidth Choices.}
		\begin{tabular}{@{}lccc@{}}
			\toprule
			&Coefficient&\multicolumn{2}{c}{Average Partial Effect}\\
			\cmidrule(lr){2-2}\cmidrule(lr){3-4}
			&ABC1&ABC1&LPM\\
			\midrule
			Participation$_{t - 1}$ & [1.580, 1.674] & [0.284, 0.303] & [0.453, 0.540] \\ 
			Middle Class & [-0.116, -0.107] & [-0.014, -0.013] & [-0.014, -0.012] \\ 
			Upper Class & [-0.347, -0.321] & [-0.044, -0.040] & [-0.043, -0.031] \\ 
			\#Children 0-1 & [-1.610, -1.594] & [-0.191, -0.190] & [-0.313, -0.303] \\ 
			\#Children 2-4 & [-0.318, -0.302] & [-0.038, -0.036] & [-0.054, -0.039] \\ 
			\#Children 5-18 & [-0.120, -0.113] & [-0.014, -0.013] & [-0.019, -0.013] \\ 
			Birth$_{t + 1}$ & [-0.520, -0.501] & [-0.066, -0.064] & [-0.082, -0.080] \\ 
			\bottomrule
		\end{tabular}
		\begin{tablenotes}
			\footnotesize
			\item\emph{Note:} The intervals show the range of the estimates; \textit{ABC1} and \textit{LPM} denote the bias-corrected estimators; bandwidths are chosen from $\{1, \ldots, 4\}$ and $\{1, \ldots, 15\}$ for \textit{ABC1} and \textit{LPM}.
			\item\emph{Further control variables:} squared age, married, east, lag of number of children between zero and one, and number of household members above 18.
			\item\emph{Source:} \textit{GSOEP} 1984--2013.
		\end{tablenotes}
	\end{threeparttable}
\end{table}

\begin{table}[!htbp]
	\centering
	\begin{threeparttable}
		\caption{\label{tab:calibrated}Results of the Calibrated Simulation Study.}
		\begin{tabular}{@{}lccccccc@{}}
			\toprule
			&\multicolumn{3}{c}{Coefficient}&\multicolumn{4}{c}{Average Partial Effect}\\
			\cmidrule(lr){2-4}\cmidrule(lr){5-8}
			&MLE&ABC1&SPJ1&MLE&ABC1&SPJ1&LPM\\
			\midrule
			\multicolumn{8}{c}{\textit{Panel A: Relative Bias}}\\
			Participation$_{t - 1}$ & -21.681 &  -7.173 & -12.310 & -38.464 & -16.452 & -24.779 &  95.404 \\ 
			Middle Class &  27.591 &   8.583 &  19.596 &  12.289 &   4.626 &  12.369 &  13.353 \\ 
			Upper Class &  24.191 &   5.734 &  31.748 &   9.089 &   1.641 &  23.081 & -10.789 \\ 
			\#Children 0-1 &  20.875 &   4.879 &  11.889 &   6.402 &   1.083 &   5.341 &  52.210 \\ 
			\#Children 2-4 &  49.423 &  10.962 &  24.895 &  31.531 &   6.946 &  19.140 &  27.993 \\ 
			\#Children 5-18 &  46.165 &  12.428 &  23.739 &  28.661 &   8.357 &  17.626 &  -5.823 \\ 
			Birth$_{t + 1}$ &  10.080 &   0.656 &   5.118 &  -4.889 &  -4.539 &  -1.693 &  23.556 \\ 
			\multicolumn{8}{c}{\textit{Panel B: Coverage Probability}}\\
			Participation$_{t - 1}$ &   0.000 &   0.000 &   0.000 &   0.000 &   0.339 &   0.000 &   0.000 \\ 
			Middle Class &   0.713 &   0.934 &   0.807 &   0.948 &   0.969 &   0.834 &   0.916 \\ 
			Upper Class &   0.884 &   0.954 &   0.483 &   0.928 &   0.933 &   0.492 &   0.930 \\ 
			\#Children 0-1 &   0.000 &   0.431 &   0.001 &   1.000 &   1.000 &   0.680 &   0.000 \\ 
			\#Children 2-4 &   0.000 &   0.647 &   0.071 &   0.348 &   0.997 &   0.195 &   0.173 \\ 
			\#Children 5-18 &   0.008 &   0.731 &   0.326 &   0.613 &   0.983 &   0.473 &   0.939 \\ 
			Birth$_{t + 1}$ &   0.714 &   0.956 &   0.877 &   0.985 &   0.983 &   0.906 &   0.333 \\ 
			\bottomrule
		\end{tabular}
		\begin{tablenotes}
			\footnotesize
			\item\emph{Note:} The biases are in percentage of the truth; \textit{MLE}, \textit{ABC1}, \textit{SPJ1}, and \textit{LPM} denote the (bias-corrected) estimators; bandwidths are 2 and 4 for \textit{ABC1} and \textit{LPM}; \textit{LPM} standard errors are robust to heteroskedasticity and clustered by woman and year; results based on 1,000 repetitions.
			\item\emph{Further control variables:} squared age, married, east, lag of number of children between zero and one, and number of household members above 18.
			\item\emph{Source:} \textit{GSOEP} 1984--2013.
		\end{tablenotes}
	\end{threeparttable}
\end{table}

\end{document}